\documentclass[11pt,reqno]{article}
\usepackage{amsmath}
\usepackage{amsfonts}
\usepackage{amssymb}
\usepackage{latexsym}
\usepackage[dvips]{graphicx}
\usepackage{epsf}

\allowdisplaybreaks \numberwithin{equation}{section}

\newcommand{\be}{\begin{equation}}
\newcommand{\ee}{\end{equation}}
\newcommand{\bi}{\begin{itemize}}
\newcommand{\ei}{\end{itemize}}
\newcommand{\bea}{\begin{eqnarray}}
\newcommand{\eea}{\end{eqnarray}}

\let\a=\alpha \let\b=\beta    
        
\let\m=\mu    \let\n=\nu

\let\==\equiv

\newcommand{\tr}{{\rm tr}}
\newcommand{\Tr}{{\rm Tr}}
\newcommand{\hh}{{\hat{h}}}

\newcommand{\cG}{\mathcal{G}}

\newcommand{\cM}{\mathcal{M}}

\newcommand{\cO}{\mathcal{O}}
\newcommand{\cP}{\mathcal{P}}
\newcommand{\cR}{\mathcal{R}}
\newcommand{\cS}{\mathcal{S}}

\newcommand{\Pt}{\tilde{\Phi}}

\newcommand{\unit}{{\bf{1}}}

\newcommand{\half}{\tfrac{1}{2}}

\newcommand{\cb}{\bar{c}}
\newcommand{\gb}{\bar{g}}

\newcommand{\I}{\mathrm{i}}
\newcommand{\e}{\mathrm{e}}
\newcommand{\p}{\partial}

\newcommand{\fb}{\bar{f}}
\newcommand{\Cb}{\bar{C}}
\newcommand{\Db}{\bar{D}}

\newcommand{\Rb}{\bar{R}}

\newcommand{\rmd}{{\rm d}}

\newcommand{\Dz}{\Delta}

\newcommand{\intg}{\int\!\! d^{d}\! x \sqrt{g}\;}
\newcommand{\intgq}{\int\!\! d^{d}\! x \sqrt{\gb}\;}
\newcommand{\bc}{\bar{c}}

\newcommand{\Pk}{\mathcal{P}_{k}}
\newcommand{\comment}[1]{}
\newcommand{\Qq}{\bar{Q}}
\newcommand{\Qt}{\tilde{Q}}
\newcommand{\Qqt}{\tilde{\bar{Q}}}
\newcommand{\Aq}{\bar{A}}
\newcommand{\At}{\tilde{A}}
\newcommand{\Aqt}{\tilde{\bar{A}}}
\newcommand{\zc}{Z^{c}_{k}}
\newcommand{\tphi}{\tilde{\Phi}}
\newcommand{\hphi}{\hat{\Phi}}
\newcommand{\cphi}{\check{\Phi}}

\begin{document}
\thispagestyle{empty}
\begin{flushright} \small
MZ-TH/10-03
\end{flushright}
\bigskip

\begin{center}
 {\LARGE\bfseries  Ghost wave-function renormalization in   \\[1.5ex]
Asymptotically Safe Quantum Gravity
}
\\[10mm]
Kai Groh and Frank Saueressig \\[3mm]
{\small\slshape
Institute of Physics, University of Mainz\\
Staudingerweg 7, D-55099 Mainz, Germany \\[1.1ex]
{\upshape\ttfamily kgroh@thep.physik.uni-mainz.de} \\
{\upshape\ttfamily saueressig@thep.physik.uni-mainz.de} }\\
\end{center}
\vspace{5mm}

\hrule\bigskip

\centerline{\bfseries Abstract} \medskip
\noindent
Motivated by Weinberg's asymptotic safety scenario, we investigate the gravitational renormalization group flow in the Einstein-Hilbert truncation supplemented by the wave-function renormalization
of the ghost fields. The latter induces non-trivial corrections to the $\beta$-functions for Newton's constant and the cosmological constant.  The resulting ghost-improved phase diagram is investigated in detail. In particular, we find a  non-trivial ultraviolet fixed point, in agreement with the asymptotic safety conjecture which also survives in the presence of extra dimensions. In four dimensions the ghost anomalous dimension at the fixed point is $\eta_c^* = -1.8$, supporting space-time being effectively two-dimensional at short distances.
\bigskip
\hrule\bigskip
\newpage
%
\section{Introduction}
Constructing a consistent and predictive quantum theory for gravity is one of the prime challenges of theoretical high energy physics today. One proposal in this direction, which recently received a lot of attention, is Weinberg's asymptotic safety scenario \cite{Weinberg:1980gg,Weinberg:2009bg,Weinberg:2009ca}, see \cite{Lauscher:2007zz,Niedermaier:2006wt,Percacci:2007sz,Reuter:2007rv,Litim:2008tt,Niedermaier:2006ns} for reviews. This scenario adopts Wilson's modern viewpoint on renormalization \cite{Wilson:1973jj}, assessing that gravity has a fundamental description within the framework of non-perturbatively renormalizable quantum field theories. The key ingredient underlying this idea is a conjectured non-trivial (or non-Gaussian) fixed point of the gravitational renormalization group (RG) flow. For RG trajectories attracted to it at high energies, the fixed point ensures that the dimensionless coupling constants remain finite, so that physical quantities are safe from unphysical UV divergences. These trajectories, also including the one describing our world \cite{Reuter:2004nx}, span the UV critical surface $\cS_{\rm UV}$ of the fixed point. The precise position of ``our RG trajectory'' within this surface is determined by free parameters, which have to be fixed by experiment. For a finite-dimensional $\cS_{\rm UV}$ this construction is then as predictive as a standard, perturbatively renormalizable, quantum field theory. Elucidating the fixed point structure of the gravitational RG flow
and understanding the properties of the corresponding UV critical surfaces is a central aspect of the asymptotic safety program to date.

An important technical tool in this program is the functional renormalization group equation (FRGE) for gravity \cite{Reuter:1996cp}. 
Formulated in terms of the Wetterich equation \cite{Wetterich:1992yh}, the FRGE describes the dependence of the effective average action $\Gamma_k$
on the coarse graining scale $k$
\be\label{FRGE}
\p_t \Gamma_k = \half {\rm STr} \left[ \left( \Gamma_k^{(2)}  + \cR_k \right)^{-1} \, \p_t \cR_k  \right]\, , \quad \p_t = k \frac{\rmd}{\rmd k} \, .
\ee
Here, $\Gamma_k^{(2)}$ denotes the second functional derivative of the effective average action with respect to the fields of the theory and {\rm STr} is a generalized functional trace which includes a minus sign for ghosts and fermions. Furthermore, $\cR_k$ is a matrix valued IR cutoff, which provides a $k$-dependent mass-term for fluctuations with momenta $p^2<k^2$. The interplay between the full regularized propagator $\left( \Gamma_k^{(2)}  + \cR_k \right)^{-1}$ and $\p_t \cR_k$ ensures that the STr receives contributions from a small $p^2$-interval around $k^2 \approx p^2$ only, rendering the trace-contribution finite.\footnote{The FRGE captures the RG-dependence of the effective action. The relation between the effective action at a fixed point $\Gamma_*$ and the corresponding fundamental action $S$ has recently been discussed in \cite{Manrique:2008zw}.}  

While \eqref{FRGE} constitutes a formally exact equation, it comes with 
the drawback that it cannot be solved exactly. A widely used approximation scheme
 for obtaining non-perturbative information from the FRGE consists of making an ansatz (truncation) for $\Gamma_k$ 
which retains a finite number of $k$-dependent coupling constants only. Projecting the 
 the resulting RG flow onto the truncation subspace allows to read off the $\beta$-functions for the running couplings  
as the coefficients of the interaction monomials retained by the ansatz without resorting to perturbation theory. 
Employing truncations, an important point is to establish the robustness of the emergent physical picture.
Within a given ansatz, its reliability may be tested by studying the dependence of physical quantities on the 
shape of the unphysical IR regulator $\cR_k$. A second, more laborious, route consists in extending the truncation
ansatz, thereby showing that all physical results, as, e.g., the fixed points of the RG flow, are robust under the extension.  

For gravity, these ideas have been implemented systematically, so far focusing on the gravitational 
sector of $\Gamma_k$ mostly. In this class the truncation ansatz is spanned by diffeomorphism invariant operators build from
the physical metric $g$, e.g., $\Gamma_k = (16 \pi G_k)^{-1} \int d^dx \sqrt{g} (-R + 2 \Lambda_k) + \ldots$. 
The most studied case, the so-called Einstein-Hilbert truncation, encompasses a scale-dependent Newton's constant $G_k$ and cosmological constant $\Lambda_k$, and has been analyzed 
in a number of works  \cite{Reuter:2001ag,Litim:2003vp,Souma:1999at,Lauscher:2001ya,Fischer:2006fz,Litim:2006dx,Fischer:2006at}.
Subsequently, this ansatz has been refined by including higher-derivative $R^2$-interactions
 \cite{Lauscher:2001rz,Lauscher:2001cq,Lauscher:2002mb}, higher order polynomials in $R$ up to $R^8$
within the framework of $f(R)$-gravity \cite{Codello:2007bd,Machado:2007ea,Codello:2008vh},
 non-local operators \cite{Reuter:2002kd,Machado:2007ea}, and lately also the Weyl-squared interactions \cite{Benedetti:2009rx,Benedetti:2009gn,Benedetti:2009iq} 
capturing the characteristic features of higher-derivative gravity. 
All these computations have identified a NGFP of the gravitational RG flow, providing substantial 
evidence for the asymptotic safety scenario. Moreover, the $f(R)$- and $C^2$-results point 
at the dimension of the associated UV critical surface being finite, possibly even as low as three.  
Notably, the essential features of this picture already emerge from the structurally significantly simpler 
flow equations obtained within the conformally reduced gravity framework \cite{Reuter:2008wj,Reuter:2008qx,Manrique:2009uh}.

In this work we go beyond the gravitational approximation of $\Gamma_k$, including quantum effects from the ghost sector.
More specifically, we will augment the Einstein-Hilbert truncation by a non-trivial wave-function renormalization of the ghost fields 
(for a similar computation in Yang-Mills theory see \cite{Ellwanger:1995qf}). Our main motivation for this ghost-improvement
originates from the analogy to QCD where this coupling plays an essential role for the IR physics of theory \cite{Alkofer:2000wg,Pawlowski:2005xe,Fischer:2006ub,Zwanziger:2003cf}.
While it is clear, that there is also a non-trivial interplay between the ghosts and the gravitational $\beta$-functions in gravity, explicit computations are not available yet. Here we close
this gap, computing the anomalous dimension of the ghost fields $\eta_c$ and its effect on the running of Newton's constant and the cosmological constant.
The explicit computation is based on a new perturbative heat-kernel technique developed by Anselmi and Benini \cite{Anselmi:2007eq} 
 which allows the systematic expansion of the flow equation in the presence of background ghost fields in a curved background.\footnote{In a companion paper \cite{Eichhorn:2010gh}, a similar analysis will be carried out using a spectrally adjusted cutoff and a flat-space projection technique. We thank A.\ Eichhorn and H.\ Gies for informing us on their upcoming work.}
This continues the exploration of quantum gravity effects in the ghost sector \cite{Eichhorn:2009ah}.

The rest of the paper is organized as follows. In Section \ref{sect:2} we derive the $\beta$-functions governing the RG dependence of Newton's constant, the cosmological constant, and the wave-function renormalization in the ghost sector, using a perturbative heat-kernel technique for non-minimal differential operators. The ghost-improved fixed point structure and phase diagram is analyzed in Section \ref{Sect:5} and we comment on our findings in Section \ref{sect:6}. Some technical details on the heat-kernel techniques and threshold functions employed in this paper are relegated to Appendix \ref{App:A}, while the functions determining the gravitational and ghost anomalous dimensions are defined in Appendix \ref{App:B}. 

\section{The ghost-improved Einstein-Hilbert truncation}
\label{sect:2}
The main purpose of this paper is the investigation of the gravitational RG flow,
including the quantum effects captured by the wave-function renormalization $Z_k^c$
in the ghost-sector. Our truncation ansatz, which will be called the ``ghost-improved Einstein-Hilbert truncation'',
 encompasses three scale-dependent coupling constants: Newton's constant $G_k$, 
 the cosmological constant $\Lambda_k$, and the power-counting marginal $Z_k^c$ multiplying the ghost-kinetic term. In this section, we will derive
  the non-perturbative $\beta$-functions capturing the RG dependence of these couplings.
\subsection{The truncation ansatz}
\label{sect:2.1}
Our ansatz for the effective average action is of the general form
\be\label{ansatz}
\Gamma_k[g,C,\Cb; \gb, c, \cb ] = 
\Gamma^{\rm grav}_k[g] + \Gamma^{\rm gf}_k[g;\gb] +
\Gamma^{\rm gh}_k[g,C,\Cb; \gb, c, \cb] \, . 
\ee
Besides on the physical metric $g$ and the classical ghost fields $C, \Cb$, it also depends on the corresponding
background fields $\gb$ and $c, \cb$. They are related by
\be\label{backgroundsplit}
g_{\m\n} = \gb_{\m\n} +h_{\m\n} \, , \qquad \Cb_\m = \cb_\m + \fb_\m \, , \qquad C_\m = c_\m + f_\m \, ,
\ee
where $h_{\m\n}$ and $f_\m, \fb_\m$ denote the expectation value of the quantum fluctuations around the background, which are not necessarily small.
In the gravitational approximation the computations are simplified by setting the background ghost fields to zero. This, however, does not allow to keep track of
the ghost-kinetic term, so that we work with a non-trivial ghost background in the following. 

The gravitational part $\Gamma^{\rm grav}_k[g]$ is build from the physical metric $g$ and taken to be of the Einstein-Hilbert form
\be
\Gamma^{\rm grav}_k[g]= 2 \, \kappa^{2} \, Z_{k}^N \intg (-R+2\Lambda_{k}) \, ,
\ee
where $\kappa^2 = (32 \pi G_0)^{-1}$ with $G_0$ a fixed reference scale and $Z_{k}^N$
denotes the wave-function renormalization for the graviton. $\Gamma^{\rm grav}_k[g]$ is 
supplemented by the gauge-fixing term $\Gamma^{\rm gf}_k[g;\gb]$. Employing the harmonic 
gauge, the latter reads
\be\label{hgauge}
\Gamma^{\rm gf}_k[h;\gb] = \half \, Z_{k}^N \, \intgq \bar{g}^{\mu\nu}F_{\mu}F_{\nu} \, , \qquad F_{\mu}=\sqrt{2}\kappa(\bar{D}^{\rho}h_{\rho\mu}-\half\bar{D}_{\mu} \gb^{\a\b} h_{\a\b}) \, .
\ee 
The resulting Faddeev-Popov determinant is captured by the ghost term
\be
\Gamma^{\rm gh}_k[g,C,\bar{C}; \gb, c, \bar{c}] = 
-\sqrt{2} \, Z^{c}_{k} \, \intgq \Cb_{\mu} \, \cM^{\mu}{}_{\nu} \, C^{\nu} \, ,
\ee
with 
\be\label{ghostkernel}
\cM^{\mu}{}_{\nu} = \gb^{\mu\rho} \gb^{\sigma\lambda} \Db_\lambda (g_{\rho \nu} D_\sigma + g_{\sigma \n} D_\rho ) - \gb^{\rho \sigma} \gb^{\m\lambda} \Db_\lambda (g_{\sigma \n} D_\rho ) \, , 
\ee
containing the wave-function renormalization of the ghosts $Z_k^c$. The gauge-choice \eqref{hgauge} has the main virtue, that it allows for a straightforward comparison to earlier results \cite{Reuter:2001ag} obtained in the Einstein-Hilbert truncation without ghost-improvement, fixing $Z_k^c = 1$.

Before entering into the explicit computation of the $\beta$-functions
arising from the ansatz \eqref{ansatz}, it is useful  
to first consider the left-hand-side of the flow equation
and identify the interaction monomials whose coefficients encode
the running of our coupling constants. Taking the $\p_t$-derivative
of our ansatz \eqref{ansatz} and setting the fluctuation fields to 
zero afterwards yields
\be\label{LHSflow}
\p_t \Gamma_k =  2 \kappa^2  \int d^dx \sqrt{\gb} \left[ - (\p_t Z^N_k) \Rb + 2 \p_t(Z^N_k\Lambda_k) \right] 
- \sqrt{2} (\p_t Z^c_k) \int d^dx \sqrt{\gb} \,  \cb^\m \Db^2 c_\m + \cdots \, . 
\ee
Thus it suffices to extract the Einstein-Hilbert monomials and the ghost
kinetic term from the right-hand-side of the flow equation. Comparing the coefficients
multiplying these interactions then allows us to read off the desired $\beta$-functions for $Z^N_k, Z^c_k$ and $\Lambda_k$. 
%
\subsection{Quadratic forms and the inverse Hessian $(\Gamma^{(2)}_k + \cR_k)^{-1}$}
\label{sect:2.2}
%
Upon specifying our truncation ansatz, we proceed by computing the second variation of $\Gamma_k$ 
 with respect to the fluctuations \eqref{backgroundsplit}. To facilitate the subsequent steps, it is convenient to decompose the metric fluctuations into
their traceless and trace part 
\be\label{Tdec}
h_{\m\n} = \hh_{\m\n} + \frac{1}{d} \gb_{\m\n} h \, , \qquad \quad h = \gb^{\m\n} h_{\m\n} \, , \qquad \gb^{\m\n} \hh_{\m\n} = 0 \, .
\ee
The forms quadratic in the fluctuations can be simplified further by first identifying $g = \gb, C = c, \Cb = \cb$ and specifying a particular class of 
backgrounds. In our case, this class has to be general enough to distinguish the interaction monomials in the Einstein-Hilbert term and the ghost-kinetic
term. This can be accomplished by choosing the background metric $\gb$ as the one of a $d$-dimensional sphere, implying
\be\label{gravback}
\Rb_{\mu\nu\rho\sigma}=\frac{1}{d(d-1)}\,\Rb\,(\gb_{\mu\rho}\gb_{\nu\sigma}-\gb_{\mu\sigma}\gb_{\nu\rho}),
\qquad \Rb_{\mu\nu}=\frac{1}{d} \, \Rb \, \gb_{\mu\nu} \, .
\ee
Moreover, the background ghost field can be taken transversal
\be\label{ghostback}
\Db_{\mu}c^{\mu}=0  \, , \qquad \Db_{\mu} \cb^{\mu}=0 \, , 
\ee
which suffices to keep track of the ghost kinetic term. In the sequel, we will resort to this choice of background to simplify all expressions.
For later reference, we also introduce the unit on the space of traceless symmetric tensors (2T) and vectors (1)
\be\label{units}
\unit_{\rm 2T} = \half \left( \delta_\m^\rho \delta_\n^\sigma + \delta_\m^\sigma \delta_\n^\rho \right) - \tfrac{1}{4} g_{\m\n} g^{\rho\sigma} \, , \qquad \unit_{\rm 1} = \delta_\m^\nu \, . 
\ee

Given these preliminaries, we now expand the ansatz \eqref{ansatz} around the background \eqref{gravback} and \eqref{ghostback}, retaining the pieces quadratic in the fluctuations only. For $\Gamma_k^{\rm grav}[g] + \Gamma_k^{\rm gf}[h,\gb]$ the result has already been given in \cite{Reuter:1996cp}
\be\label{grav:quad}
\Gamma^{\rm quad}_k[h; \gb] = \half \kappa^2 Z^N_k \intgq \left\{
\hh_{\m\n} \left[ \Dz - 2 \Lambda_k + C_T \Rb \right] \hh^{\m\n} - \tfrac{d-2}{2d} h \left[ \Dz - 2 \Lambda_k + C_S \Rb \right] h
\right\}
\ee
with
\be
C_T = \frac{d^2-3d+4}{d(d-1)} \, , \qquad C_S = \frac{d-4}{d} \, , 
\ee
and $\Dz \equiv - \Db^2$ denoting the covariant Laplacian constructed from the background metric.
Owed to the non-trivial background ghost field, the analogous computation for $\Gamma_k^{\rm gh}$ is slightly more involved.
Thus we first give the intermediate results obtained from expanding \eqref{ghostkernel} contracted with a dummy vector $C^\nu$
in $h_{\m\n}$
\be\label{Mexp}
\begin{split}
\overline{\cM}^{\, \mu}{}_{\nu} \, C^\nu = & \, \left[ \Db^2 + \tfrac{1}{d} \, \Rb \right] \, C^\mu \, , \\
\overline{\delta_h \cM} ^{\, \mu}{}_{\nu} \, C^\nu = & \, \gb^{\m\rho} \gb^{\sigma \lambda} \Db_\lambda \left[ h_{\rho\alpha} \Db_\sigma + h_{\sigma \alpha} \Db_\rho + (\Db_\alpha h_{\rho \sigma} ) \right] C^\alpha \\
& - \gb^{\rho\sigma} \gb^{\mu\lambda} \Db_\lambda \left[ h_{\sigma \alpha} \Db_\rho + \half (\Db_\alpha h_{\rho \sigma} )  \right] C^\alpha \, , \\
\overline{\delta_h^2 \cM} ^{\,\mu}{}_{\nu} \, C^\nu = &  \, 0 \, . \\
\end{split}
\ee
Here the bar indicates that we expand around $g = \gb$. Remarkably, the last line is independent of the choice of background or gauge. Based on these results, we obtain the quadratic form in the ghost sector
\be\label{ghost:quad}
\begin{split}
\Gamma_k^{\rm quad,gh} & =
	 \sqrt{2} Z^c_k \intgq \Big\{
	\bar{f}_\m \left[ \Dz - \tfrac{1}{d} \Rb \right] f^\m -
	h A^\a \bar{f}_\a - \hh^{\m\n} Q_{\m\n}{}^{\a} \bar{f}_\a -
	h \Aq_\a f^\a - \hh_{\m\n} \Qq^{\m\n}{}_{\a} f^\a
	\Big\} \\
& =  \sqrt{2} Z^c_k \intgq \Big\{
	\bar{f}_\m \left[ \Dz - \tfrac{1}{d} \Rb \right] f^\m -
	\bar{f}_\a \At^\a h - \bar{f}_\a \Qt^{\a}{}_{\m\n} \hh^{\m\n} -
	f^\a \Aqt_\a h - f^\a \Qqt_{\a}{}^{\m\n} \hh_{\m\n}
	\Big\} \, .
\end{split}
\ee
Here the two lines are equivalent up to surface terms and define the Grassmann-valued operators $A, \bar{A}, Q, \bar{Q}$, and their adjoints, respectively.
Their explicit expressions read
\be\label{ghkernels}
\begin{split}
\At^\a = & \, \tfrac{1}{d}\left[
	\Db^{2}c^{\alpha}+\Rb^{\alpha\sigma}c_{\sigma}+\Db_{\sigma}c^{\alpha}\Db^{\sigma}+(2-\tfrac{d}{2})\Db^{\alpha}c^{\sigma}\Db_{\sigma} 
	+(1-\tfrac{d}{2})c^{\sigma}\Db^{\alpha}\Db_{\sigma}\right], \\
\Aq_\alpha = & \, - \tfrac{1}{d} \left[ \Db^\sigma \bar{c}_\a \Db_\sigma + \Db_\a \bar{c}^\sigma \Db_\sigma \right] \, ,  \\
A^\a = & \tfrac{1}{d} \left[ \Db_\sigma c^\a \Db^\sigma + \Db^\a c^\sigma \Db_\sigma - (1-\tfrac{d}{2} ) c^\sigma \Db_\sigma \Db^\alpha \right] \, , \\
\Aqt_\a = & - \tfrac{1}{d} \left[
	\Db^2 \bar{c}_\a + \Db^\sigma \bar{c}_\a \Db_\sigma + \Rb_{\a \sigma} \bar{c}^\sigma + \Db_\a \bar{c}^\sigma \Db_\sigma \right] ,
\end{split}
\ee
and
\be
\begin{split}
\Qt^{\alpha}{}_{\mu\nu} =& \,
	\delta^{\alpha}_{(\mu}\Db^{2}c_{\nu)} + R_{(\mu}{}^\a{}_{\n)}{}^\sigma c_\sigma + \delta^{\alpha}_{(\mu}c^{\sigma}\Db_{\nu)}\Db_{\sigma} \\
	& + \delta^{\alpha}_{(\mu}\Db^{\sigma}c_{\nu)}\Db_{\sigma} + \Db^{\alpha}c_{(\nu}\Db_{\mu)} + \delta^{\alpha}_{(\mu}\Db_{\nu)}c^{\sigma}\Db_{\sigma} - \Db_{(\mu}c_{\nu)}\Db^{\alpha} \\
	& - \frac{1}{d}\gb_{\mu\nu} ( \Db^{2}c^{\alpha} + \Rb^{\alpha\sigma}c_{\sigma} + \Db^{\sigma}c^{\alpha}\Db_{\sigma}
	+ 2\Db^{\alpha}c_{\sigma}\Db^{\sigma} + c^\sigma \Db^\a \Db_\sigma ) , \\
\Qq^{\m\n}{}_\a = & \,
	\Db_{\alpha}\Db^{(\nu}\bc^{\mu)} - \delta_{\alpha}^{(\nu}\Db_{\sigma}\bc^{\mu)}\Db^{\sigma}
	 - \delta_{\alpha}^{(\nu} \Db^{\mu)}\bc_{\sigma}\Db^{\sigma} + \Db^{(\nu}\bc^{\mu)}\Db_{\alpha} \\
	& + \frac{1}{d}\gb^{\mu\nu} ( \Db^{\sigma}\bc_{\alpha}\Db_{\sigma} + \Db_{\alpha}\bc_{\sigma}\Db^{\sigma} ) , \\
Q_{\m\n}{}^\a = & \, \delta_{(\m}^\a \Db^\sigma c_{\n)} \Db_\sigma - c_\sigma \Db^\sigma \Db_{(\nu} \delta_{\m)}^\a + \Db^\a c_{(\mu} \Db_{\n)} - \Db_{(\m} c_{\n)} \Db^\a \\
	& \quad - \tfrac{1}{d} \gb_{\m\n} \left[ - c_\sigma \Db^\sigma \Db^\a + \Db^\sigma c^\a \Db_\sigma + \Db^\a c^\sigma \Db_\sigma \right] \, , \\
\Qqt_\a{}^{\m\n} = & \, - \delta^{(\m}_\a \left[
	\Db_\sigma \bar{c}^{\n)} \Db^\sigma + \Db^2 \bar{c}^{\nu)} + \Db^{\n)} \bar{c}_\sigma \Db^\sigma + \Rb^{\n)\sigma} \bar{c}_\sigma  
	\right] + \Db^{(\m} \bar{c}^{\nu)} \Db_\alpha \\
	& \quad + \tfrac{1}{d} \gb^{\m\n} \left[
	\Db^\sigma \bar{c}_\a \Db_\sigma + \Db^2 \bar{c}_\a + \Db_\a \bar{c}_\sigma \Db^\sigma + \Rb_{\a\sigma} \bar{c}^\sigma
	\right] ,
\end{split}
\ee
where the covariant derivatives to the left of the ghost fields act on $c$ or $\bar{c}$ only and $(\m\n) = \half (\m\n + \n\m)$ denotes symmetrization with unit strength. The last lines in the $Q$-expressions ensure that the operators are traceless in the 2T-indices $\m\n$.
 The quadratic forms \eqref{grav:quad} and \eqref{ghost:quad} provide the key ingredients for constructing the IR cutoff $\cR_k$ and the inverse Hessian $(\Gamma^{(2)}_k+\cR_k)^{-1}$ in the sequel. 

The construction of the IR-cutoff follows the prescription in \cite{Reuter:1996cp}. I.e., at the level of the path integral, the gauge-fixed action is supplemented by
an IR regulator of the form
\be\label{IRcut}
\Delta_kS = \int d^dx \sqrt{\gb} \left\{ \half \, \hh_{\m\n} \, \cR_{k,\hh\hh}^{\m\n\a\b} \, \hh_{\a\b} + \half \, h \, \cR_{k,hh} \, h + \fb^\mu \, \cR_{k,{\fb f}} \, f_\nu  
\right\} \, . 
\ee
The matrix $\cR_k$ is designed in such a way, that it adds a $k$-dependent mass-term to the Laplacians appearing in the kinetic terms,
\be\label{IRrule}
\Dz \rightarrow P_k \equiv \Dz + R_k \, , 
\ee  
where $R_k = k^2 R^{(0)}_k(\Dz/k^2)$ is the scalar part of the IR regulator and $R^{(0)}_k$ a dimensionless shape-function interpolating monotonically between $R^{(0)}_k(0) = 1$ and $\lim_{z \rightarrow \infty} R^{(0)}_k(z) = 0$. Comparing \eqref{IRcut} to the quadratic forms \eqref{grav:quad} and \eqref{ghost:quad}, this prescription fixes 
\be
\cR_{k,\hh\hh} = \unit_{\rm 2T} \, \kappa^2 \, Z^N_k \, R_k 
\, , \quad 
\cR_{k,hh} = - \tfrac{d-2}{2d} \, \kappa^2 \, Z^N_k \, R_k 
\, , \quad 
\cR_{k,\fb f} = \sqrt{2} \, \unit_{\rm 1} \, Z^c_k \, R_k \, .
\ee
Observe that $\cR_k$ inherits a non-trivial $k$-dependence via the wave-function renormalization factors. Notably, it is this feature which is partially responsible for the non-perturbative character of the computation.

The next step consists in constructing the Hessian $\Gamma^{(2)}_k$ and the inverse of $\Gamma_k^{(2)} + \cR_k$. Here we follow the conventions
 \cite{Pawlowski:2005xe,Ellwanger:1995qf}, using a skew-symmetric metric to couple 
anti-commuting fields (see Appendix A of \cite{Pawlowski:2005xe} for details).\footnote{One can explicitly verify that the resulting FRGE is equivalent to \cite{Lauscher:2001ya}.} In terms of the multiplets
\be
\Phi = \{ \hh_{\m\n} , h , f^\a , \fb_\a \} \, , \qquad \bar{\Phi} = \{ \hh_{\m\n}, h, - \fb_\a, f^\a \} \, ,
\ee
the Hessian is given by the following matrix
\be
\left[ \Gamma^{(2)}_k \right]^{ij}(x, y) = \frac{1}{\sqrt{\gb(x)} \sqrt{\gb(y)}} \, \frac{\delta^2 \Gamma_k}{\delta 
\bar{\Phi}_i(x) \delta \Phi_j(y)} \, ,
\ee
where all variations act from the left.
Substituting the quadratic forms \eqref{grav:quad} and \eqref{ghost:quad}, 
\be\label{prop}
\left( \Gamma_k^{(2)} + \cR_k \right)^{ij} = \left[
\begin{array}{cc}
	K & Q \\
	\tilde{Q} & M \\
\end{array} \right]
\ee
assumes block form with entries
\be
\begin{split}
K=& \, \kappa^{2} \, Z_{k}^N \, {\rm diag} \left[ (P_k -2\Lambda_{k} +C_T \Rb) \unit_{\rm 2T} \, , \,  - \tfrac{d-2}{2d}(P_k - 2\Lambda_{k} + C_S \Rb) \right] \, , \\
M = & \, \sqrt{2} \, Z^{c}_{k} \, (P_k -\tfrac{1}{d} \Rb) \, {\rm diag} \left[ \, \unit_1 \, , \, \unit_1 \, \right] \, , 
\end{split}
\ee
and
\be
Q =  \sqrt{2} Z^c_k \left[
\begin{array}{cc}
\Qq^{\m\n}{}_\a & Q^{\m\n,\a} \\
\Aq_\a & A^\a
\end{array} \right] 
\, , \qquad
\tilde{Q} = \sqrt{2} Z^c_k \left[
\begin{array}{cc}
\Qt^{\a,\m\n} & \At^\a \\
-\Qqt_\a{}^{\m\n} & -\Aqt_\a
\end{array} \right] \, .
\ee

The inverse of \eqref{prop} is then found via the general inversion formula for $2\times 2$-block matrices
\be\label{Ginv}
\left[\begin{array}{cc} K & Q \\ \tilde{Q} & M \end{array}\right]^{-1} = \left[\begin{array}{cc} \left(K - Q M^{-1} \tilde{Q}\right)^{-1} & - K^{-1}Q\left(M - \tilde{Q} K^{-1} Q \right)^{-1} \\ -M^{-1} \tilde{Q} \left(K - Q M^{-1} \tilde{Q} \right)^{-1} & \left(M - \tilde{Q} K^{-1} Q\right)^{-1} \end{array}\right]\,.
\ee
Expanding the inverse up to second order in the background-ghost fields, and taking into account the minus sign originating from the ghost sector of the supertrace, the right-hand-side of the flow equation  becomes
\be\label{flow2}
\begin{split}
\p_t \Gamma_k  = & \, \half \Tr_{\rm gr} \left[ (K^{-1} + K^{-1} Q M^{-1} \tilde{Q} K^{-1} ) \p_t \cR_k^{\rm grav} \right] \\
& \quad - \half \Tr_{\rm gh}  \left[ (M^{-1} + M^{-1} \tilde{Q} K^{-1} Q M^{-1} ) \p_t \cR_k^{\rm gh} \right] \, , \\
 := & \; \cS_{\rm 2T} + \cS_0 + \cS_1 + \cG_{\rm 2T} + \cG_0 + \cG_1 + \cdots \, ,
\end{split}
\ee
with $\cR_k^{\rm grav} = {\rm diag}[\cR_{k,\hh\hh}, \cR_{k,hh} ]$, $\cR_k^{\rm gh} = {\rm diag}[\cR_{k,\fb f},\cR_{k,\fb f}]$, and
terms not contributing to the truncation indicated by the dots. As already anticipated in the last line, the full trace decomposes into operator traces on the space of 
traceless symmetric tensors (2T), scalars (0) and vectors (1). Substituting the explicit expressions for the block matrices, the traces $\cS_i$, which by definition, do not include $Q$ and $\tilde{Q}$ and are thus independent of the background ghost fields, become
\be\label{Strace}
\begin{split}
\cS_{\rm 2T} = & \, \tfrac{1}{2} \Tr_{\rm 2T} \left[ \frac{\unit_{\rm 2T}}{ Z^N_k (P_k - 2 \Lambda_k + C_T \Rb)} \, \p_t (Z^N_k \, R_k) \right] \, , \\ 
\cS_{\rm 0} = & \, \tfrac{1}{2} \Tr_{\rm 0} \left[ \frac{1}{Z^N_k (P_k - 2 \Lambda_k + C_S \Rb)} \, \p_t (Z^N_k \, R_k) \right] \, ,  \\
\cS_{\rm 1} = & \, -  \Tr_{\rm 1} \left[ \frac{\unit_1}{ Z^c_k (P_k - \tfrac{1}{d} \Rb)} \, \p_t (Z^c_k \, R_k) \right] \, .
\end{split}
\ee
They give rise to the $\beta$-functions for Newton's constant and the cosmological constant.
The $\beta$-function for $Z^c_k$ is captured by the terms of second order in the background-ghost fields. Neglecting the curvature terms and making use of the cyclicity of the trace,
these are found as
\be\label{Gtrace}
\begin{split}
\cG_{\rm 2T} = & - \frac{Z^c_k}{\sqrt{2} \kappa^2 (Z^N_k)^2} \Tr_{\rm 2T} \left[
	\frac{\p_t(Z^N_k R_k)}{[P_k - 2 \Lambda]^2 } \left(
	Q^{\m\n}{}_\a \frac{1}{P_k} \Qqt^\a{}_{\rho\sigma} - \Qq^{\m\n}{}_\a \frac{1}{P_k} \Qt^{\a}{}_{\rho\sigma}
	\right) \right] \, , \\
\cG_{0} = &  \frac{2d}{d-2} \frac{Z^c_k}{\sqrt{2} \kappa^2 (Z^N_k)^2} \Tr_{0} \left[
	\frac{\p_t(Z^N_k R_k)}{[P_k - 2 \Lambda]^2 }  \left(
	A^\a \frac{1}{P_k} \Aqt_\a - \Aq_\alpha \frac{1}{P_k} \At^\a
	\right) \right] \, , \\
\cG_{1} = & - \frac{1}{\sqrt{2} \kappa^2 Z^N_k} \Tr_{1} \bigg[
	\frac{\p_t(Z^c_k R_k)}{P_k^2} \bigg(
	\Qt_{\a}{}^{\m\n} \frac{1}{P_k - 2 \Lambda} \Qq_{\m\n}{}^{\b} - \Qqt^{\m\n}{}_\a \frac{1}{P_k - 2 \Lambda} Q_{\m\n}{}^\beta \\
&	- \frac{2d}{d-2} \Big( \At_{\a} \frac{1}{P_k - 2 \Lambda} \Aq^{\b} - \Aqt_{\a} \frac{1}{P_k - 2 \Lambda} A^{\b} \Big) \bigg) \bigg] \, .
\end{split}
\ee
Notably, the insertions including the background ghosts, like, e.g., the $Q P^{-1} \Qqt$ and $-\Qq P^{-1} \Qt$ in $\cG_{\rm 2T}$, always appear in pairs.
Each part thereby gives exactly the same contribution to the $\cG_i$, consistent with the symmetry factors of the underlying Feynman-diagrams. Evaluating
both parts individually provides a highly non-trivial crosscheck when computing the coefficients of the ghost-kinetic term in
the next subsection.
%
\subsection{The ghost-improved $\beta$-functions}
\label{sect:2.3}
We are now in the position to construct the $\beta$-functions for
$Z^N_k, Z^c_k$, and $\Lambda_k$. In this context, it is useful to introduce the anomalous dimensions
 of Newton's constant and the ghost wave-function renormalization 
\be
\eta_N = - \p_t \ln(Z^N_k) \, , \qquad  \eta_c = - \p_t \ln(Z^c_k) \, ,
\ee
together with the dimensionless couplings
\be
g_k = G_k k^{d-2} = (Z_k^N)^{-1} \, G_0 \, k^{d-2} \, , \qquad \lambda_k = \Lambda_k k^{-2} \, .
\ee

As explained in Section \ref{sect:2.1}, the $\beta$-functions for Newton's and the cosmological constant are encoded at zeroth order in the background-ghost fields.
The desired interaction monomials are generated by the traces $\cS_i$, eq.\ \eqref{Strace}, which are evaluated straightforwardly 
by applying the early-time heat-kernel expansion detailed in Appendix \ref{A.1} followed by inserting the identity \eqref{PhiQrel1}.
Equating the resulting coefficients with \eqref{LHSflow} provides the flow equations
for $\Lambda_k$ and $G_k = (32 \pi Z^N_k \kappa^2)^{-1}$
\be\label{beta1}
\begin{split}
\p_t \left( \tfrac{\Lambda_k}{8 \pi G_k} \right) = & \tfrac{k^d}{(4 \pi)^{d/2}} \left\{
 \tfrac{ d (d+1)}{2} \left[ \Phi^{1,0}_{d/2}(-2\lambda_k) - \half \eta_N \Pt^{1,0}_{d/2}(-2\lambda_k) \right] 
- 2 d \left[ \Phi^{1,0}_{d/2}(0) - \half \eta_c \Pt^{1,0}_{d/2}(0) \right] 
\right\} \, , \\
- \p_t \left( \tfrac{1}{16 \pi G_k} \right) = & \tfrac{k^{d-2}}{(4 \pi)^{d/2}} \Big\{
\tfrac{d(d+1)}{12} \left[  \Phi^{1,0}_{d/2-1}(-2\lambda_k) - \half \eta_N \Pt^{1,0}_{d/2-1}(-2 \lambda_k) \right] \\ & \qquad \quad
- \tfrac{d(d-1)}{2} \left[ \Phi^{2,0}_{d/2}(-2\lambda_k) - \half \eta_N \Pt^{2,0}_{d/2}(-2 \lambda_k) \right] \\ & \qquad \quad
- \tfrac{d}{3} \left[  \Phi^{1,0}_{d/2-1}(0) - \half \eta_c \Pt^{1,0}_{d/2-1}(0) \right] 
- 2 \left[ \Phi^{2,0}_{d/2}(0) - \half \eta_c \Pt^{2,0}_{d/2}(0) \right] 
\Big\} \,.  \\
\end{split}
\ee
 In the limit $Z^c_k = 1, \eta_c = 0$, this result agrees with earlier computations \cite{Reuter:1996cp,Reuter:2001ag}. The terms proportional to $\eta_c$ are novel
and capture the backreaction of the quantum effects in the ghost sector on the running of the gravitational coupling constants.

The final step is the computation of $\eta_c$. This requires extracting the background-ghost kinetic term
from \eqref{Gtrace}. Unfortunately, the differential operators entering into these traces are not minimal, so that the early-time heat-kernel expansion
is no longer applicable and a more sophisticated method is needed.
Here we follow Anselmi and Benini \cite{Anselmi:2007eq}, which allows to compute operator traces of Laplace-type operators which include arbitrary (non-Laplace type) ``perturbations'' 
by considering the heat-kernel expansion at non-coincident points. Adapting these techniques to the case where the perturbation is build from the background-ghost fields, 
this method becomes applicable to $\cG_i$. (See Appendix \ref{A.2} for more technical details.)  
Retaining the ghost-kinetic term only, a tedious but straightforward computation yields
\be\label{Gtrace2}
\begin{split}
\cG_{\rm 2T} = & - \tfrac{\sqrt{2}}{(4 \pi)^{d/2}} \tfrac{Z^c_k}{ \kappa^2 (Z^N_k)^2}
	\Big[ \tfrac{4d^2-d-8}{4d} Q_{d/2+1}[W_1^{N}] - \tfrac{d^2-2}{2d} Q_{d/2+2}[W_2^{N}] \Big] \, \int d^dx \sqrt{\gb} \, \cb^\m \Db^2 c_\m \, , \\
\cG_0 = & - \tfrac{\sqrt{2} }{(4 \pi)^{d/2}}  \tfrac{Z^c_k}{ \kappa^2 (Z^N_k)^2}
	\Big[ \tfrac{d-4}{d(d-2)} Q_{d/2+1}[W_1^{N}] - \tfrac{1}{d} Q_{d/2+2}[W_2^{N}] \Big] \int d^dx \sqrt{\gb} \, \cb^\m \Db^2 c_\m , \\
\cG_1 = & \, - \tfrac{1}{(4 \pi)^{d/2}} \tfrac{1}{\sqrt{2} \kappa^2 Z^N_k}
	\Big[ \tfrac{2 d^2- 5 d-2}{2(d-2)} Q_{d/2+1}[W_1^{c}] + d \, Q_{d/2+2}[W_2^{c}]  \Big] \int d^dx \sqrt{\gb} \, \cb^\m \Db^2 c_\m \, .
\end{split}
\ee
Here the $Q$-functionals are defined in \eqref{Qfct} and 
\be\label{ffct}
\begin{array}{ll}
W_1^{N} \equiv \frac{\p_t(Z^N_k R_k)}{(\Pk - 2 \Lambda)^2} \, \frac{1}{\Pk} \, , \qquad &
W_2^{N} \equiv - \frac{\p_t(Z^N_k R_k)}{(\Pk - 2 \Lambda)^2} \left. \frac{\rmd}{\rmd x} \frac{1}{\cP_k(x)} \right|_{x = \Dz} \, , \\[1.3ex]
W_1^{c} \equiv   \frac{\p_t(Z^c_k R_k)}{\Pk^2} \, \frac{1}{\Pk - 2 \Lambda} \, , \qquad &
W_2^{c} \equiv - \frac{\p_t(Z^c_k R_k)}{\Pk^2} \left. \frac{\rmd}{\rmd x} \frac{1}{\cP_k(x)- 2 \Lambda} \right|_{x = \Dz} \, .
\end{array}
\ee
Substituting \eqref{PhiQrel2} and equating the result with the ghost kinetic term in \eqref{LHSflow},
we finally find
\be\label{dtZc}
\begin{split}
\p_t \zc = \frac{4 \, \zc \, g_k}{(4\pi)^{d/2-1}} \bigg\{ & \,
C_{\rm gr} \left( \Phi^{2,1}_{d/2+1}(-2 \lambda_k) - \half \eta_N \tphi^{2,1}_{d/2+1}(-2\lambda_k) \right) \\
& \, + C_{\rm gh} \left( \Phi^{1,2}_{d/2+1}(-2 \lambda_k) - \half \eta_c \tphi^{1,2}_{d/2+1}(-2 \lambda_k) \right) \\
& \, + d \left( \eta_N - \eta_c \right) \left( \tphi^{2,2}_{d/2+2}(-2\lambda_k) + \hphi^{2,2}_{d/2+2}(-2\lambda_k) \right)
\bigg\} \, , 
\end{split}
\ee
where
\be\label{Ccdef}
C_{\rm gr} = \frac{4d^2-9d-2}{d-2} \, , \qquad C_{\rm gh} = \frac{2d^2-5d-2}{d-2} \, . 
\ee
Here the $\Phi^{2,2}_{d/2+2}$ and $\cphi^{2,2}_{d/2+2}$-terms originating from $Q_{d/2+2}[W_2^I]$ drop out, due to the cancellation of the corresponding coefficients.

The $\beta$-functions are then obtained by solving eqs.\ \eqref{beta1} and \eqref{dtZc} for $\p_t \lambda_k$, $\p_t g_k$ and $\eta_c$. This yields
\be\label{bfcts}
\p_t \lambda_k = \beta_\lambda \, , \qquad \p_t g_k = \beta_g = (d-2+\eta_N) \, g_k \, , 
\ee
with
\be\label{blambda}
\begin{split}
\beta_\lambda = & \, - (2 - \eta_N) \lambda_k + \half g_k (4\pi)^{1 - d/2} \\ 
& \quad \left[  2 d(d+1) \Phi^{1,0}_{d/2}(-2\lambda_k)  - 8 d \Phi^{1,0}_{d/2}(0) - d(d+1) \eta_N \Pt^{1,0}_{d/2}(-2\lambda_k)
 + 4 d  \eta_c \Pt^{1,0}_{d/2}(0) \right] \, ,
\end{split}
\ee
and the anomalous dimensions
\be\label{etaI}
\begin{split}
\eta_N = & \, \frac{g B_1(\lambda) +g^2 \Big[C_3(\lambda)C_4(\lambda)-B_1(\lambda)C_2(\lambda)\Big] }
{1 -g \Big[B_2(\lambda)+C_2(\lambda)\Big] +g^2 \Big[B_2(\lambda)C_2(\lambda)-C_1(\lambda)C_3(\lambda)\Big] } \, , \\
\eta_c = & \, \frac{g C_4(\lambda) +g^2 \Big[B_1(\lambda)C_1(\lambda)-B_2(\lambda)C_4(\lambda)\Big] }
{1 -g \Big[B_2(\lambda)+C_2(\lambda)\Big] +g^2 \Big[B_2(\lambda)C_2(\lambda)-C_1(\lambda)C_3(\lambda)\Big] } \, . \\
\end{split}
\ee
The explicit form of the functions $B_i(\lambda)$ and $C_i(\lambda)$ is given in Appendix \ref{App:B}. The eqs.\ \eqref{blambda} and \eqref{etaI} are the desired $\beta$-functions governing the RG dependence of $g_k, \lambda_k$, and $\eta_c$ and constitute the central result of this section.

Some comments are now in order. The inclusion of the wave-function renormalization for the ghosts gives non-trivial contributions to the $\beta$-functions for $g, \lambda$. These encompass the terms proportional to $\eta_c$ in $\beta_\lambda$ and the qualitatively new $g^2$-terms in $\eta_N$. The results obtained within the standard Einstein-Hilbert truncation \cite{Reuter:1996cp,Reuter:2007rv} are recovered by setting $C_i = 0$. Comparing powers of $g$, the leading contributions from the ghost sector are suppressed by one power of $g$, relative to the leading Einstein-Hilbert terms. Thus in the classical regime, $g \ll 1$, the ghost-improvement may be neglected. In the quantum regime close to the NGFP where $g \approx 1$, however, we expect that these corrections become important. 
Investigating the influence of these new terms on the gravitational RG flow is then the subject of the next section.

\section{Properties of the flow equation}
\label{Sect:5}
After deriving the $\beta$-functions \eqref{bfcts} and \eqref{etaI}, we now proceed
by studying their properties, mostly resorting to numerical methods. In this context,
it is useful to observe that $Z^c_k$ enters into $\beta_\lambda$ via
$\eta_c$ only and is in turn completely determined by $g_k, \lambda_k$. Thus,
substituting the explicit formula for $\eta_c$ into $\beta_\lambda$, the running of $Z^c_k$ decouples
and allows to analyze the gravitational RG flow in the two-dimensional $g,\lambda$-subsystem.
Once a RG trajectory for $g_k, \lambda_k$ is found, it can be plugged back 
into $\eta_c$ to obtain the running of the ghost anomalous dimension. We will now exploit 
this decoupling and first discuss the fixed point structure of the ghost-improved Einstein-Hilbert truncation
for  $d=4$ in Section \ref{sect5.1}, before focusing on the phase portrait and the fixed point structure including
 extra-dimensions in Sections \ref{sect5.2} and \ref{sect5.3}, respectively. 
\subsection{Fixed points of the four-dimensional $\beta$-functions}
\label{sect5.1}
As highlighted in the introduction, the crucial ingredient of the asymptotic
safety scenario is the fixed point structure of the gravitational $\beta$-functions.
Thus, we start our investigation by looking for fixed points $g^*, \lambda^*$ where $\beta_g = \beta_\lambda = 0$  
simultaneously. In the vicinity of such a point,
the linearized $\beta$-functions are given by $\p_t g_i = {\bf B}_{ij} (g_j - g_j^*)$, where
${\bf B}_{ij} = \p_{g_j} \beta_{g_i}|_{g = g^*}$, $g_i = \{g,\lambda\}$.
The stability coefficients $\theta_i$, defined as minus the eigenvalues of ${\bf B}_{ij}$, provide
an important characteristic of the fixed point. In particular, eigendirections with a positive (negative)
real part $\theta$ are UV-attractive (UV-repulsive) for trajectories close to the fixed point. For the remainder of this subsection, we will set $d=4$.

Inspection of \eqref{bfcts} immediately gives the Gaussian fixed point (GFP)
\be\label{GFP}
g^* = 0 \, , \qquad \lambda^* = 0 \, , \qquad \eta_N^* = \eta_c^* = 0 \, .
\ee
This fixed point corresponds to the free theory, and constitutes a saddle point in the $g$-$\lambda$-plane. It has one
attractive and one repulsive eigendirection with stability coefficients given by the canonical mass dimensions of 
$G$ and $\Lambda$, respectively.

The numerical analysis of the ghost-improved $\beta$-functions also reveals a unique NGFP. Its position and properties
are shown in the first two lines of Table \ref{t.NGFP}. 
\begin{table}
\begin{center}
\begin{tabular}{|c|c|c|c|c|c|c|c|} \hline
Truncation & $\lambda^*$ & $g^*$     & $g^* \lambda^*$ & $\eta_c^*$ & Re$(\theta)$ & Im$(\theta)$ &  \; cutoff  \; \\ \hline
EH + ghost & $0.135$     &  $0.859$  & $0.116$         & $-1.774$   &  \; $1.935$  \;      &  \; $2.012$  \;      & opt \\
EH + ghost & $0.260$     &  $0.355$  & $0.092$         & $-1.846$   & $2.070$      & $2.439$      & exp ($s=1$) \\ \hline
EH         & \; $0.193$  \; &  \;  $0.707$  \; &  \; $0.136$  \; &  \; $-$          & $1.475$      & $3.043$      & opt \\ \hline
\end{tabular}
\caption{Properties of the NGFP arising from the ghost-improved $\beta$-functions \eqref{bfcts}. The first two lines show the position, the universal product $g^* \lambda^*$, the ghost anomalous dimension $\eta_c^*$, and the stability coefficients of the fixed point obtained with the optimized cutoff \eqref{Ropt} and exponential cutoff \eqref{Rexp} with $s=1$, respectively. For comparison, the third line displays the characteristics of the NGFP found in the standard Einstein-Hilbert trunaction \cite{Codello:2008vh}.}
  \label{t.NGFP}
\end{center}
\end{table}
It is situated at $g^*>0, \lambda^*>0$ and UV-attractive in both $g, \lambda$. Substituting its position into $\eta_c$ determines the ghost anomalous
dimension $\eta_c^* = -1.8$.\footnote{When including the marginal $Z^c_k$ in the set of coupling constants, $\eta_c^*$ is also the stability coefficient associated with the new 
(UV-irrelevant) eigendirection. Here we refrain from adopting this viewpoint, however, since the running of $\eta_c$ is completely determined by $g_k, \lambda_k$ so that $Z^c_k$ is, most likely, not an essential coupling.} 
For comparison, the third line of Table \ref{t.NGFP} displays the properties of the NGFP obtained within the standard Einstein-Hilbert truncation {\it without}
ghost-improvement. We observe that the actual numerical values of the physical product $g^* \lambda^*$ and the stability coefficients 
are shifted by approximately $30\%$. This is in the typical range for the cutoff-scheme dependence observed in \cite{Codello:2008vh}.
Most remarkably, both the standard and the ghost-improved Einstein-Hilbert truncation give rise to the same fixed point structure. This is highly non-trivial, as
the new contributions to the gravitational $\beta$-functions are of the same order of magnitude as the other, already known, terms.  
We interpret this result as a striking confirmation of the gravitational fixed point structure disclosed by the standard Einstein-Hilbert truncation \cite{Souma:1999at,Lauscher:2001ya,Reuter:2001ag,Litim:2003vp,Codello:2008vh}. 

\begin{figure}[t]
  \centering
    \includegraphics[width=0.42\textwidth]{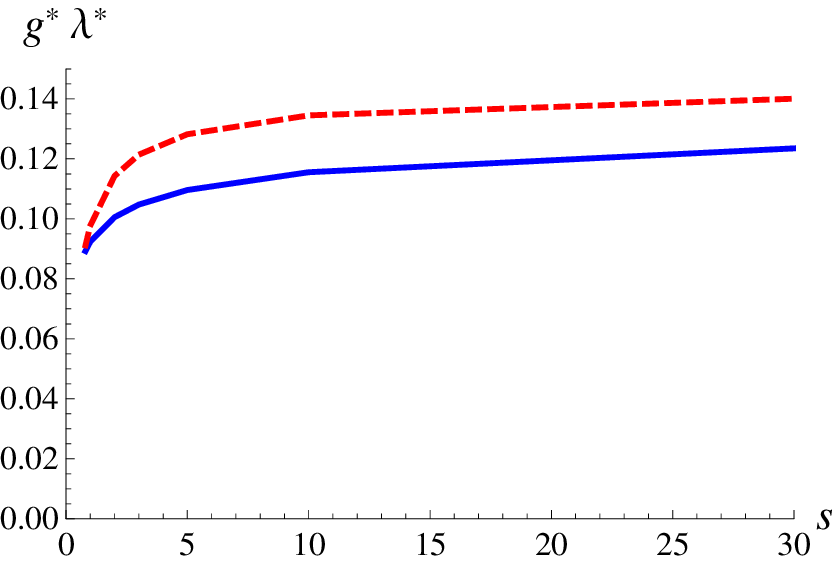} \;\;\;
    \includegraphics[width=0.42\textwidth]{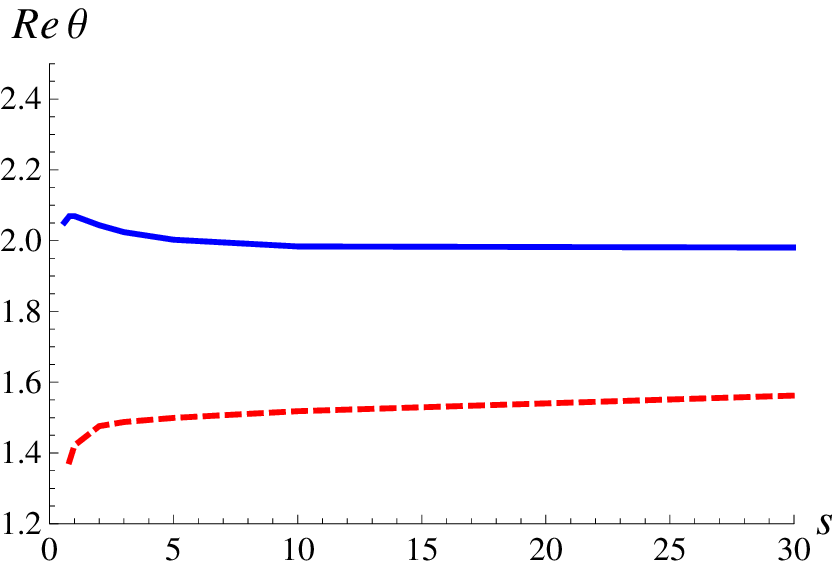} \\
    \includegraphics[width=0.42\textwidth]{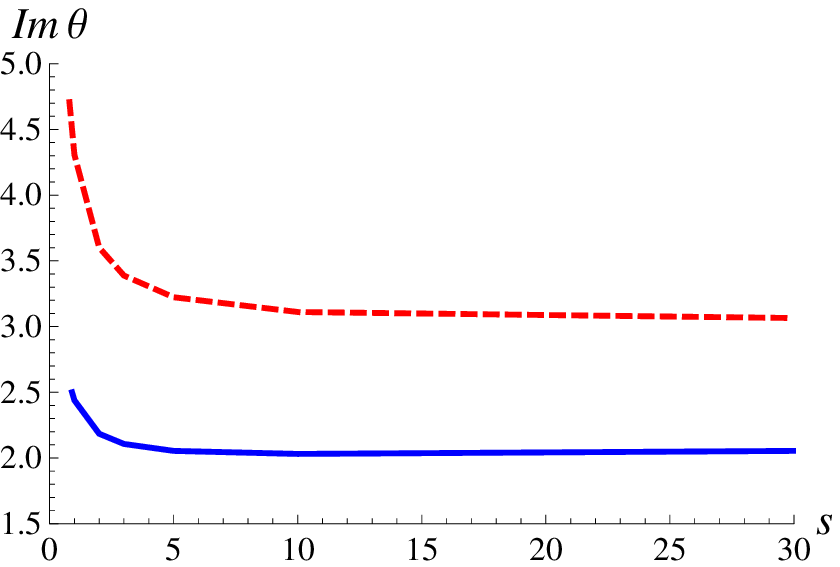} \;\;\;
    \includegraphics[width=0.42\textwidth]{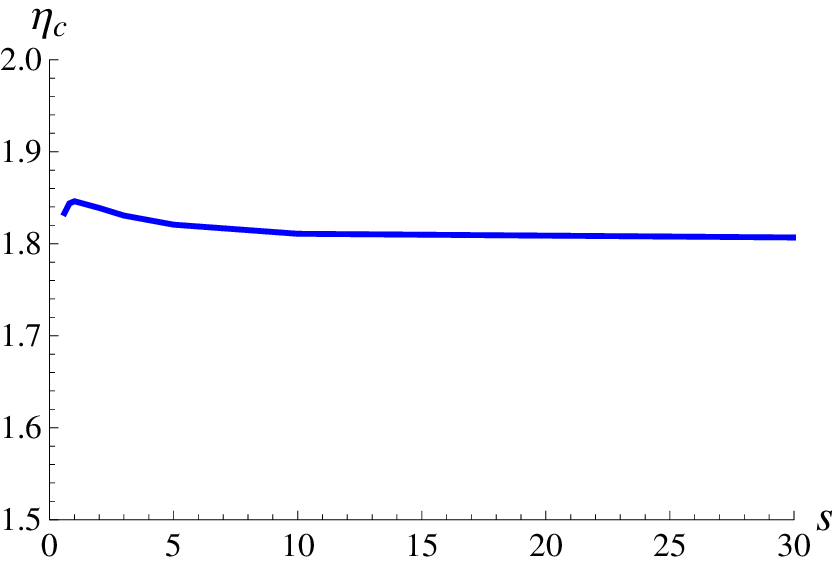}
   \caption{Stability analysis for universal quantities in the Einstein-Hilbert truncation (dashed red line), previously obtained in \cite{Reuter:2001ag}, and upon including the ghost wave-function renormalization (solid blue line). The ghost-improvement \emph{significantly decreases} the cutoff-scheme dependence.}
  \label{p.stab}
\end{figure}

The main virtue of the ghost-improvement becomes apparent, when investigating the 
stability of the physical quantities $g^* \lambda^*$, $\theta$, $\eta_c^*$ with respect to the variation of the IR cutoff $R_k$. 
To illustrate this point, we resort to the exponential cutoff \eqref{Rexp}, and determine the properties of the NGFP for varying shape parameter $s$.
Fig.\ \ref{p.stab} shows the resulting cutoff-scheme dependence of the physical quantities for the standard (dashed red line) and ghost-improved
(solid blue line) computation, respectively.
Remarkably, the ghost-improvement {\it reduces} the unphysical cutoff-scheme dependence by factors $1.4$, $2.0$ and $3.2$ for the product $g^* \lambda^*$, Re$\theta$ and 
Im$\theta$, respectively. The scheme-dependence of $\eta_c^*$ can only be determined in the ghost-improved computation. Here the variation is approximately $2\%$.
Since the cutoff-scheme dependence provides the prototypical probe for judging the reliability of a truncation, these results clearly
indicate that the ghost-improvement {\it significantly improves} the quality of the Einstein-Hilbert approximation.
In particular the fixed point properties given in Table \ref{t.NGFP} should be more robust as the ones obtained without ghost-improvement. 
%

\subsection{The ghost-improved phase portrait}
\label{sect5.2}
After analyzing the fixed point structure, we determine the phase portrait resulting from the ghost-improved $\beta$-functions. We start by investigating
the gravitational RG flow on the $g$-$\lambda-$plane, before studying the behavior of the anomalous dimensions along some typical sample trajectories.

\begin{figure}[t]
  \centering
    \includegraphics[width=0.9\textwidth]{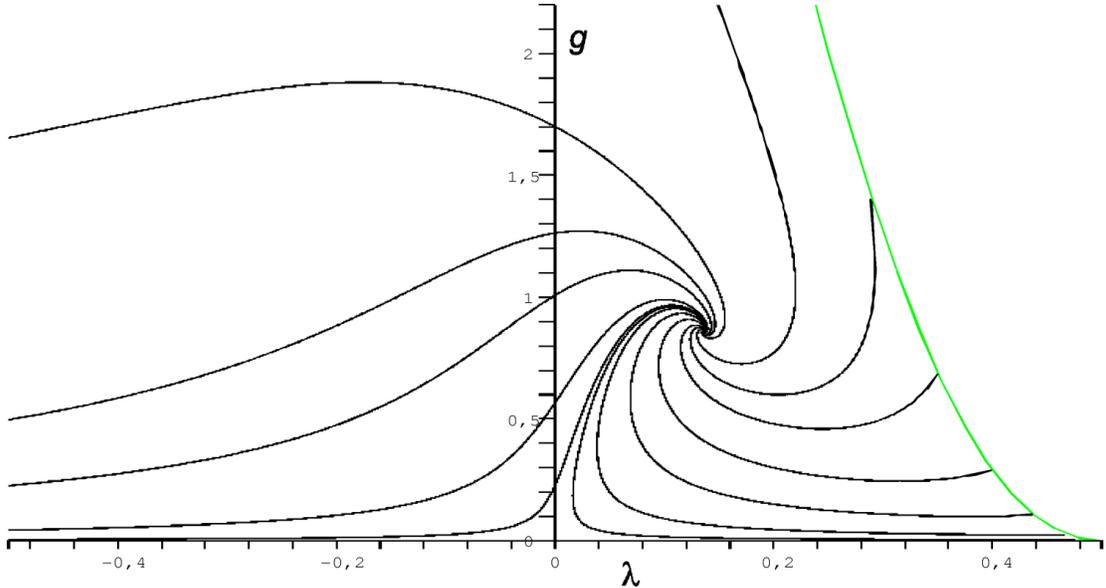}
   \caption{The RG flow of $g_k$, $\lambda_k$ obtained from the numerical integration of the ghost-improved $\beta$-functions \eqref{bfcts} using the optimized cutoff. The green line indicates a boundary of the coupling-constant space where the $\beta$-functions diverge. The phase portrait is in complete agreement with the one obtained from the standard Einstein-Hilbert truncation \cite{Reuter:2001ag,Codello:2008vh,Litim:2008tt}.}
  \label{p.phase}
\end{figure}

The phase portrait resulting from the numerical integration of the $\beta$-functions \eqref{bfcts} is depicted in Fig.\ \ref{p.phase}.
We first observe that $g_k = 0$ is a fixed line, which cannot be crossed by the flow. 
For $g_k > 0$ the flow is dominated by the interplay of the NGFP and the GFP. In this regime, the UV behavior of the RG trajectories is controlled by the NGFP, which acts as a UV attractor. Following the RG flow from this fixed point towards the IR, the RG trajectories undergo a crossover from the NGFP to the ``classical regime'' dominated by the GFP.
 Depending on whether the trajectory turns to the left (Type Ia), right (Type IIIa) or hits the GFP (Type IIa), the classical theory has a negative, positive or zero cosmological constant. The trajectories with positive cosmological constants, however, cannot be continued to $k =0$, but terminate at a finite value of $k$ when reaching the boundary of the phase space. The latter is indicated by the green line, which constitutes a singularity in the $\beta$-functions at finite $g, \lambda$. Observe that the ghost-improved phase portrait, Fig.\ \ref{p.phase}, is in complete qualitative agreement with the standard Einstein-Hilbert truncation \cite{Reuter:2001ag,Codello:2008vh,Litim:2008tt}.

\begin{figure}[t]
  \centering
    \includegraphics[height=42mm,width=0.43\textwidth]{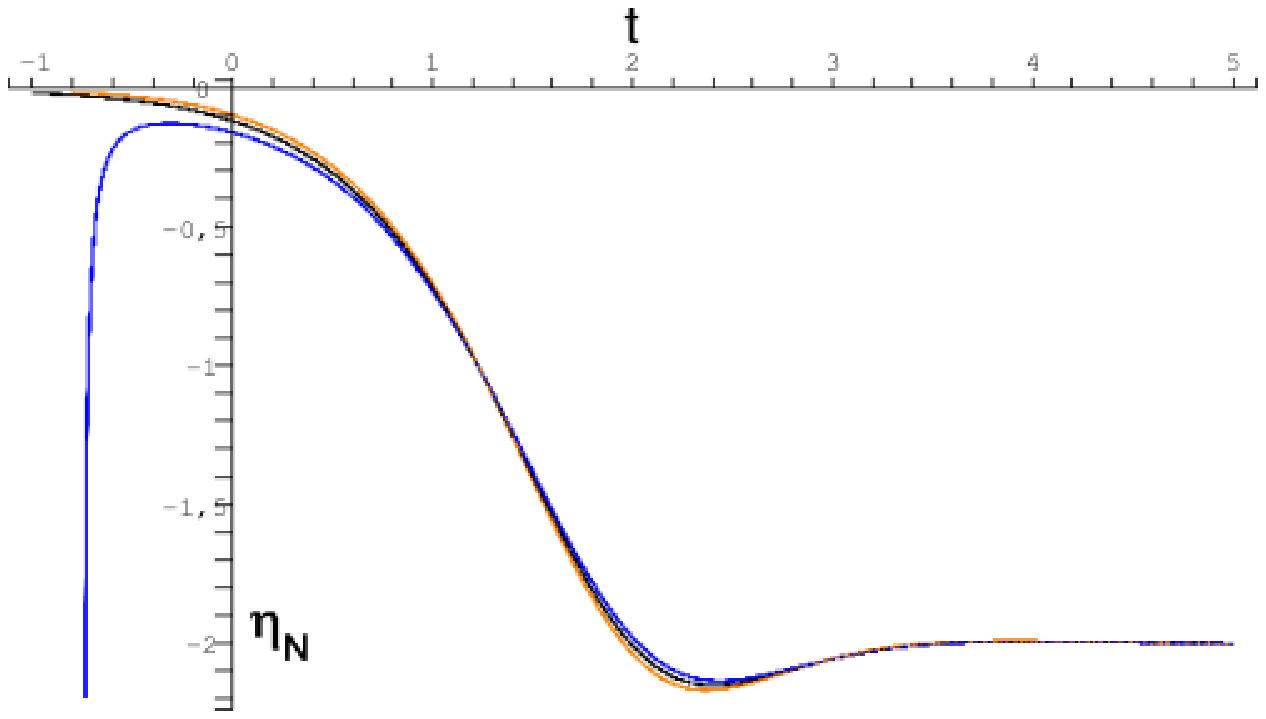} \;\;\;
    \includegraphics[height=42mm,width=0.43\textwidth]{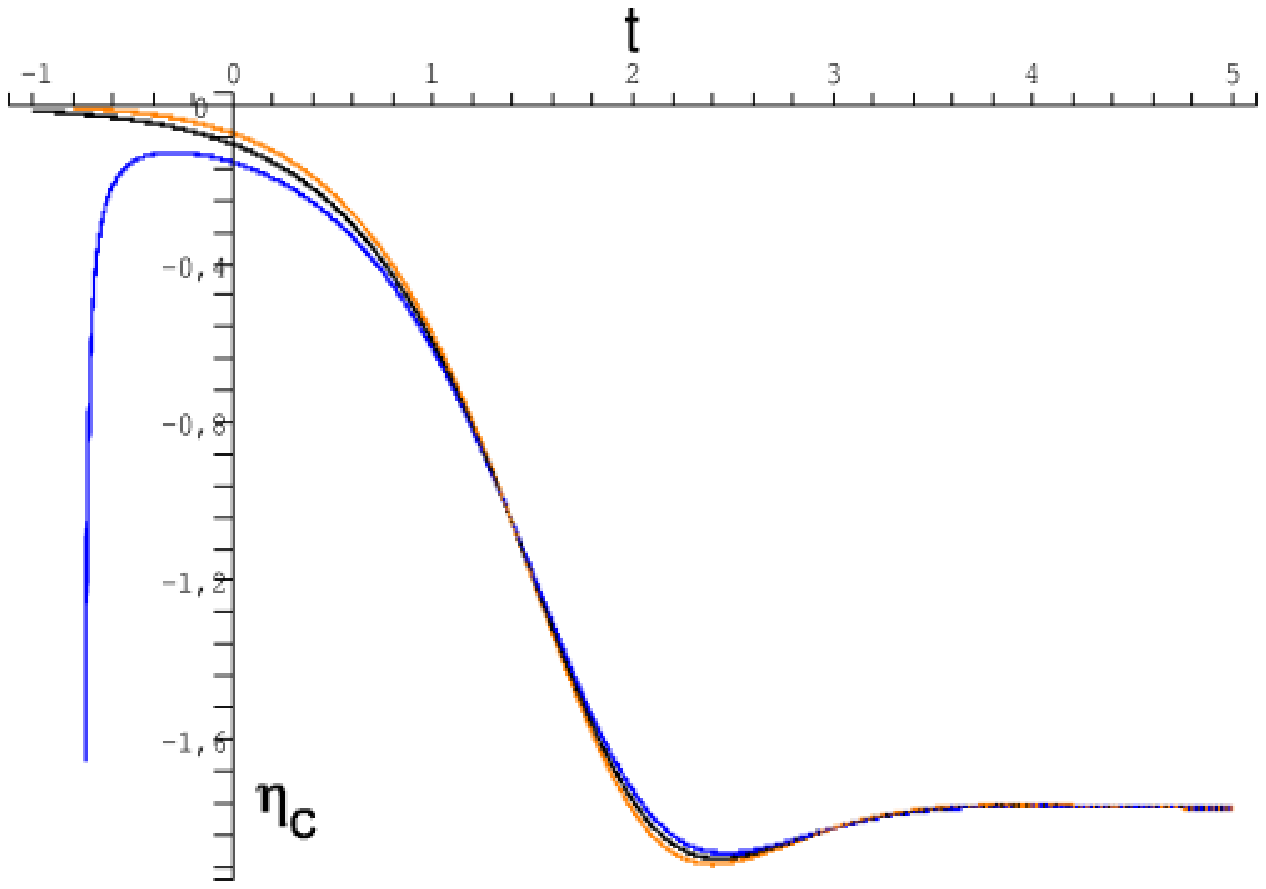} \\[1.2ex]
    \includegraphics[height=42mm,width=0.43\textwidth]{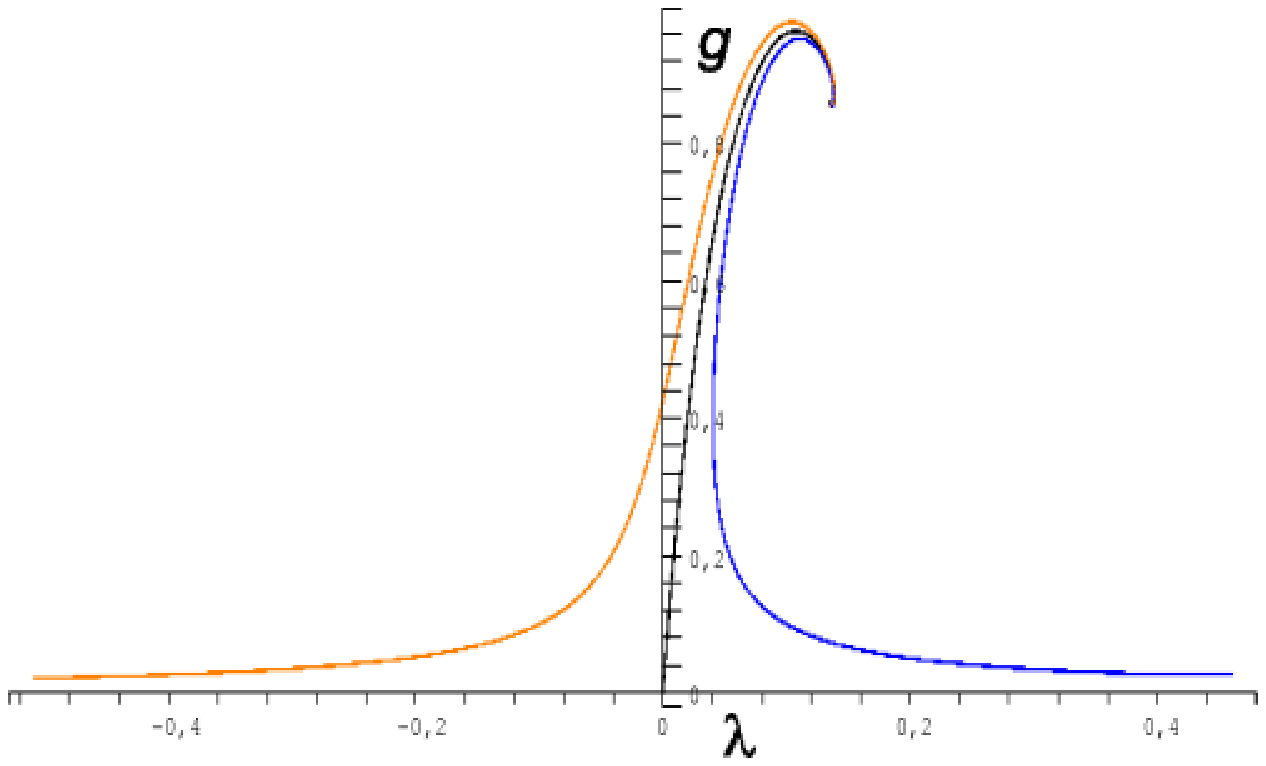} \;\;\;
    \includegraphics[height=42mm,width=0.43\textwidth]{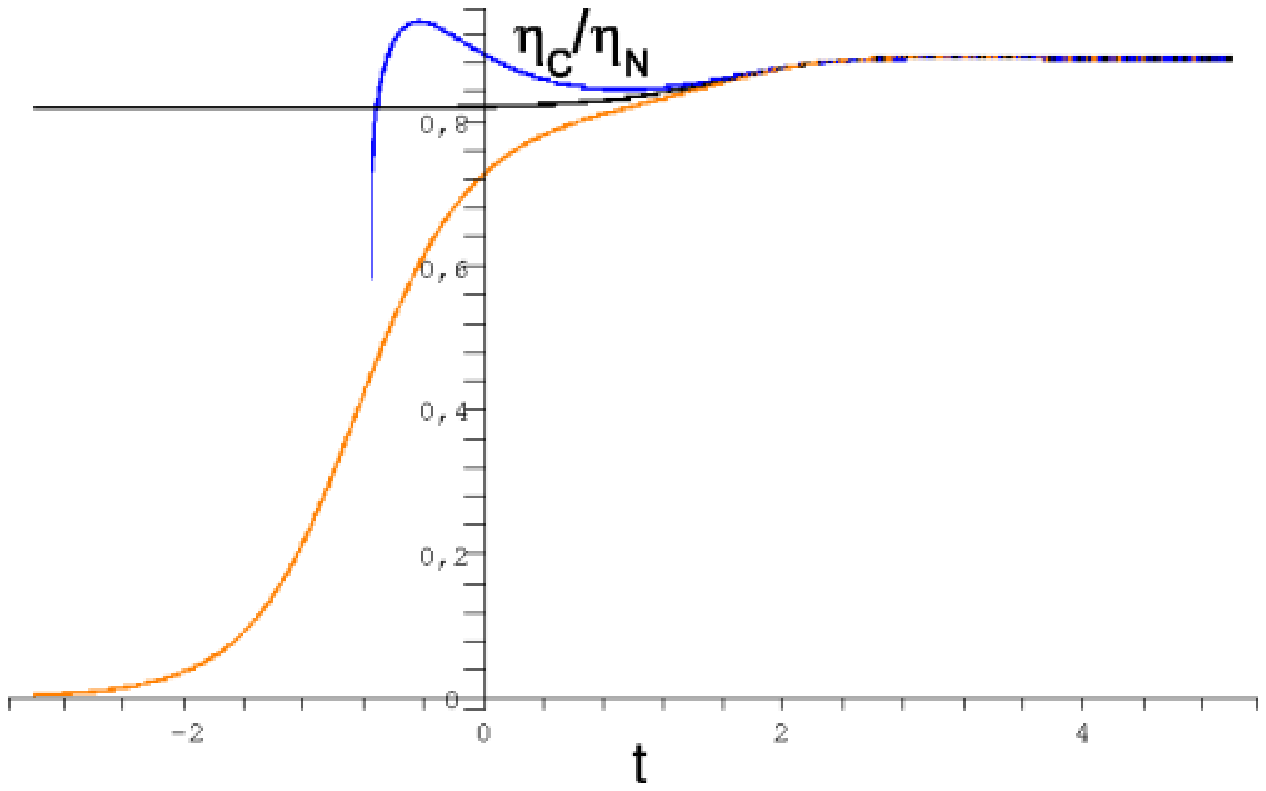}
   \caption{RG flow of the anomalous dimensions $\eta_N$ (upper left), $\eta_c$ (upper right), and their ratio $\eta_c/\eta_N$ (lower right) along the sample RG-trajectories of Type Ia (orange), Type IIa (black), and Type IIIa (blue) given in the lower left diagram.}
  \label{p.sample}
\end{figure}

It is now illustrative to pick one sample trajectory for each of the classes discussed above 
and study the $k$-dependence of the anomalous dimensions $\eta_N$ and $\eta_c$ along the flow.
This is shown in Fig.\ \ref{p.sample}. 
In the UV, for large values of $t = \ln(k/k_0)$, the anomalous dimensions are determined by the NGFP, so that $\eta_N^* = -2$ and $\eta_c^* = -1.77$. 
 Lowering $t$ and approaching the IR, the anomalous dimensions undergo a crossover towards the classical theory with $\eta_N \approx 0$, $\eta_c \approx 0$. 
 The steep increase at the end of the Type IIIa trajectory is caused by the singularity of the $\beta$-functions (green line in Figure \ref{p.phase}), and heralds the termination of the trajectory at a finite value $k$. The UV-limit of the ratio $\eta_c / \eta_N$ is governed by the NGFP and takes the value $\eta_c^* / \eta_N^* = 0.89$. Following the flow towards the IR, the ratio undergoes a crossover and asymptotes to $0$, the cutoff-scheme dependent value $54 \Phi^{0,3}_3/(24 \Phi^{0,2}_2 - \Phi^{0,1}_1)$, or a finite value at the termination point, for trajectories of Type Ia, Type IIa, and Type IIIa, respectively.

It is worthwhile to have a closer look at the singularity causing the termination of the Type IIIa trajectories. Fig.\ \ref{p.sample} shows, that at this point in coupling constant space, the anomalous dimensions $\eta_N$ and $\eta_c$ diverge. This can be traced back to the vanishing of the denominators in \eqref{etaI}. Here, the ghost-improvement has, however, a very non-trivial effect: while the denominator arising from the standard Einstein-Hilbert truncation has a term linear in $g$ only, the ghost-improvement adds an additional piece quadratic in $g$. This may provide an elegant mechanism for lifting this singularity by shifting the zeros of the denominator to complex values $g$. However, for the $B_i, C_i$ given in Appendix \ref{App:B}, this mechanism is not realized, and may requires a further improvement of the truncation before becoming operational.

\subsection{The NGFP in space-times with extra-dimensions}
\label{sect5.3}
The $\beta$-functions \eqref{blambda} and \eqref{etaI} are
continuous in the space-time dimension. In the sequel,
we will exploit this feature and analyze the resulting
gravitational fixed point structure for general $d$.
Based on the standard Einstein-Hilbert truncation,
a similar analysis has already been carried out in 
 \cite{Reuter:2001ag,Fischer:2006fz,Fischer:2006at}.
Motivated by the recent interest in asymptotically safe TeV-scale gravity models
 \cite{Litim:2007iu,Koch:2007yt,Gerwick:2009zx,Burschil:2009va},
which hinge on the existence of the NGFP in the presence 
of extra-dimensions, it is worthwhile to complement
the previous results by including the ghost-improvements. 
 
\begin{figure}[t]
  \centering
    \includegraphics[width=0.3\textwidth]{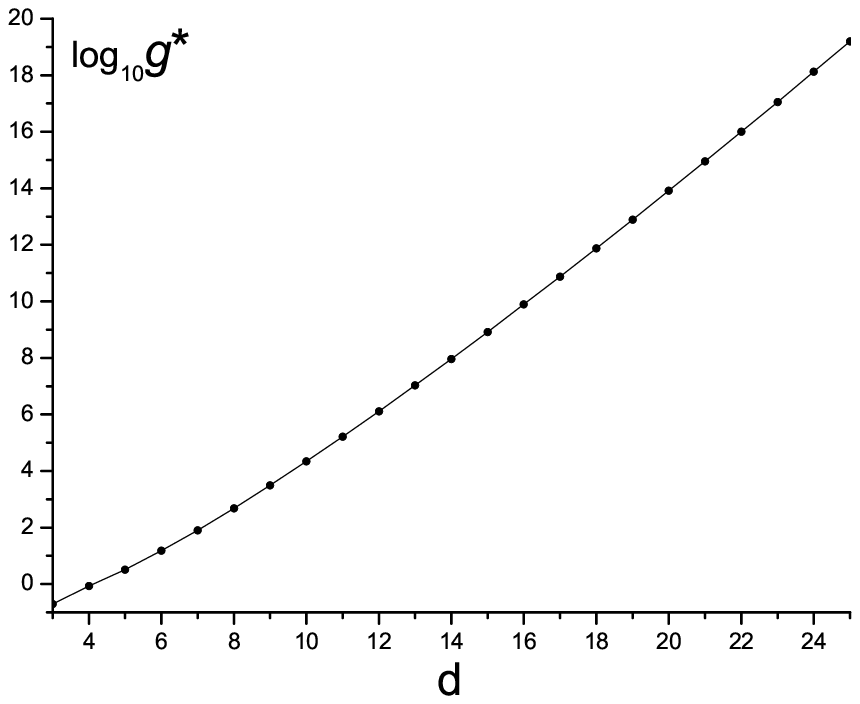} \;
    \includegraphics[width=0.3\textwidth]{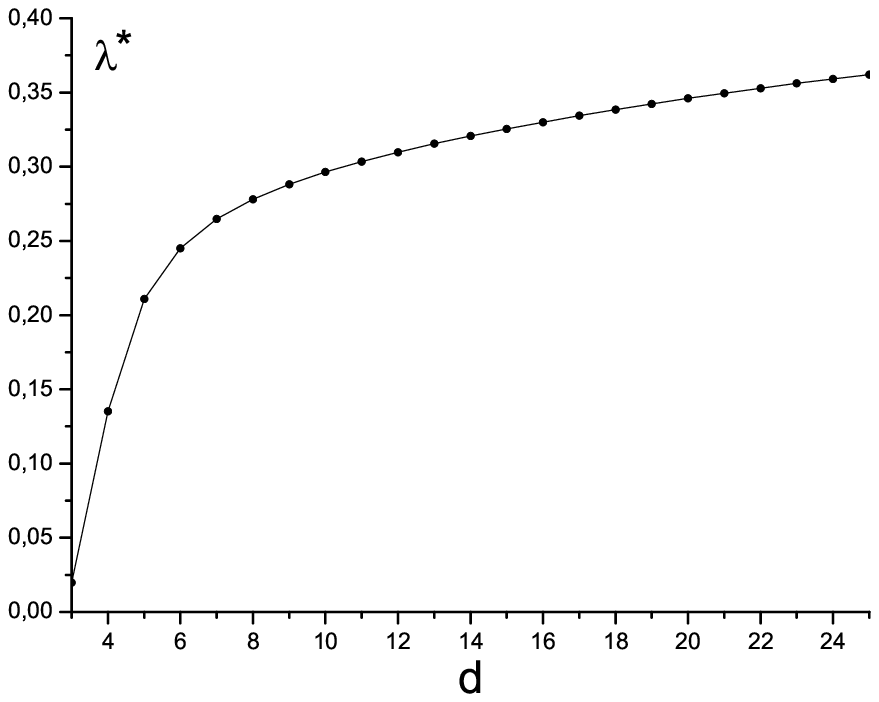} \;
    \includegraphics[width=0.3\textwidth]{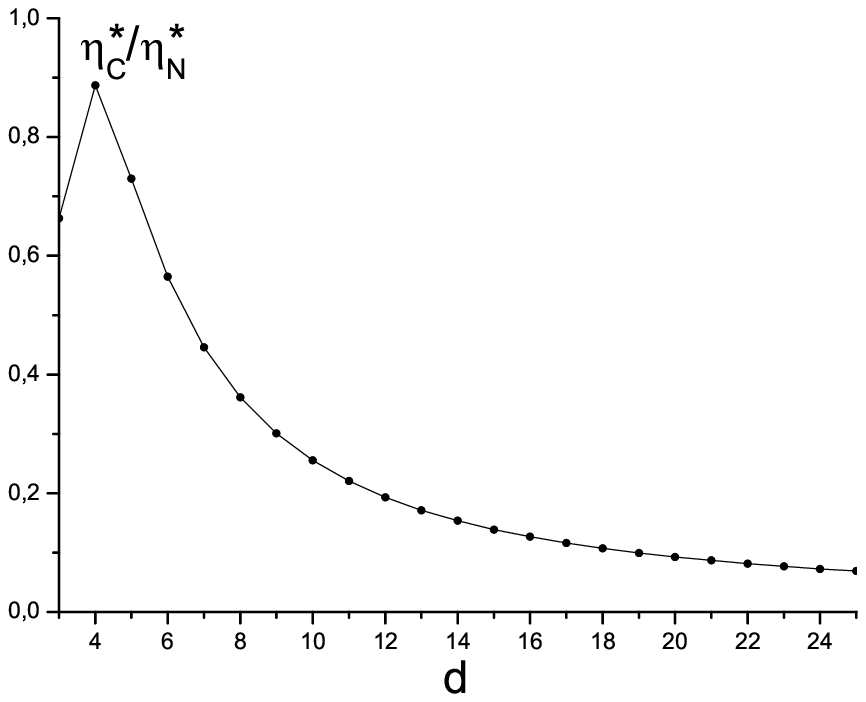} \; \\
    \begin{center}
    \includegraphics[width=0.3\textwidth]{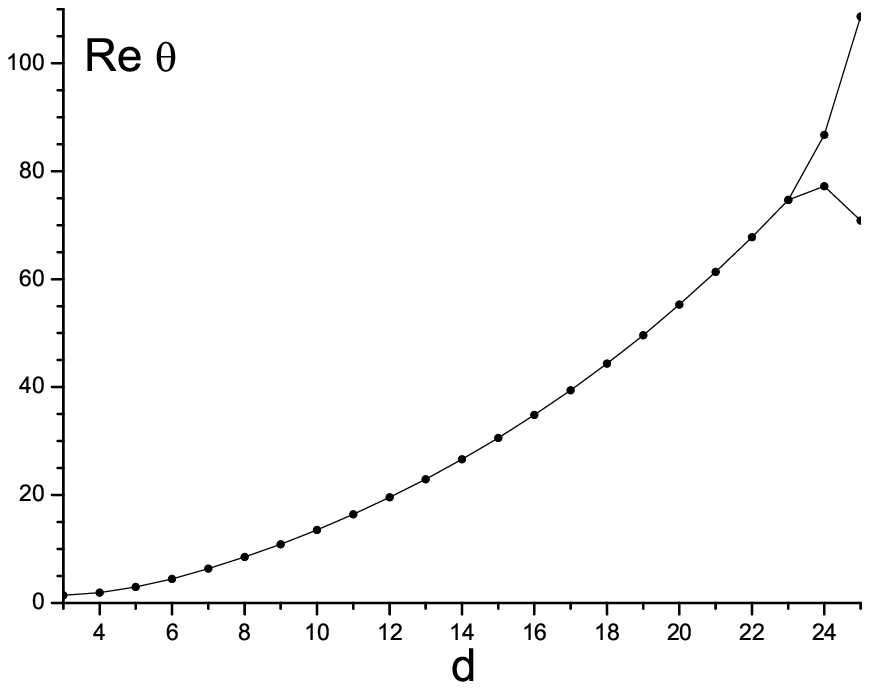} \;
    \includegraphics[width=0.3\textwidth]{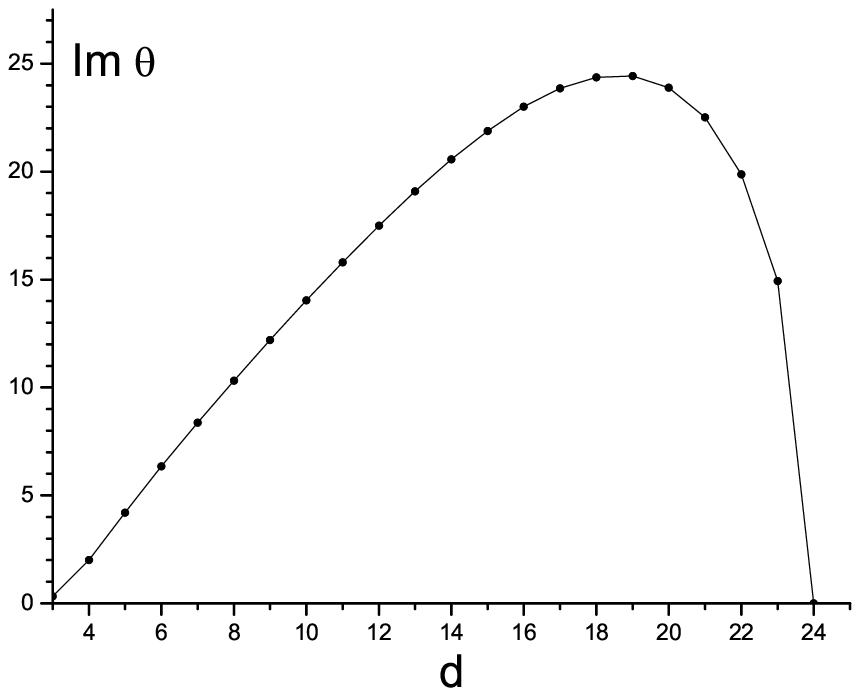} \;
    \end{center}
   \caption{Properties of the NGFP for general space-time dimension $d$, obtained with the optimized cutoff \eqref{Ropt}. There is a unique NGFP for all $3 \le d \le 25$ considered.}
  \label{p.extra}
\end{figure}

Remarkably, the $d$-dimensional fixed point structure is
strikingly similar to the one obtained in four dimensions. 
The GFP \eqref{GFP} exists for all $d$. Furthermore,
there is a unique generalization of the NGFP
for all dimensions $3 \le d \le 25$ considered here.
Its properties are summarized in Fig.\ \ref{p.extra}. 
The fixed point is situated at positive $g^* > 0$, $\lambda^* > 0$ and UV attractive
in both $g$ and $\lambda$. Below $d < 24$ its critical exponents are given by a complex pair Re$\theta \pm \I$Im$\theta$. For $d \ge 24$ the imaginary part of the critical exponents vanishes and we have two real stability coefficients, which are still UV attractive. All these results are in perfect agreement with earlier findings based on the standard Einstein-Hilbert truncation. An interesting feature arises in the ratio $\eta_c^*/\eta_N^*$, with $\eta_N^* = 2-d$, shown in the top right diagram of Fig.\ \ref{p.extra}. This ratio is peaked at $d=4$, where it reaches almost unity. For both $d > 4$ and $d < 4$ the ratio decreases. 
It is a rather curious observation that in $d=4$ the graviton and ghost propagators have the same anomalous dimension in the UV, highlighting this particular value of the space-time dimension.\footnote{A related argument, concluding that $d=4$ is special, is based on the spectral dimension of space-time computed from  the running of Newton's constant and the cosmological constant and has been put forward in \cite{Lauscher:2005qz}.} While it is clearly desirable to get a better understanding of this result, the underlying analysis is beyond the scope of the present paper.

\section{Discussion and Conclusions}
\label{sect:6}
In this paper we have analyzed the gravitational renormalization group flow
arising from the ghost-improved Einstein-Hilbert truncation. Besides a scale-dependent 
Newton's constant and cosmological constant, this setup also takes into account
the quantum effects captured by the non-trivial wave-function renormalization
in the ghost sector. 
Our main result is the occurrence of a unique non-Gaussian fixed point
of the gravitational $\beta$-functions. It is located at positive Newton's constant and cosmological constant
 and UV attractive in both $g$ and $\lambda$. Its properties are strikingly similar to 
the non-Gaussian fixed point found in the standard Einstein-Hilbert truncation \cite{Reuter:2001ag,Litim:2003vp,Souma:1999at,Lauscher:2001ya,Fischer:2006fz,Litim:2006dx,Fischer:2006at}.
This finding further substantiates the asymptotic safety scenario of quantum gravity \cite{Weinberg:1980gg}.\footnote{Qualitatively, our picture is also confirmed by the very recent results \cite{Eichhorn:2010gh}, which study the ghost-improved Einstein-Hilbert truncation employing a spectrally adjusted cutoff. The numerical variations observed in the two computations are within the typical range expected from the different cutoff-schemes \cite{Codello:2008vh}.} The fixed point also persists for higher-dimensional space-times, 
 solidifying the foundations for asymptotically safe TeV-scale gravity models \cite{Litim:2007iu,Koch:2007yt,Gerwick:2009zx,Burschil:2009va},  
possibly accessible at the LHC .

The main virtue of the ghost-improvement  
is a significant decrease of the unphysical
cutoff-scheme dependence in physical observables as, e.g., the critical exponents of the  
fixed point or the universal product $g^* \lambda^*$. Since this scheme dependence
constitutes a standard test for the quality of a truncation, we conclude that the ghost-improvement significantly increases the robustness
of the emerging physical picture. In particular, the characteristics of the fixed point displayed in Table \ref{t.NGFP}
 should provide a better approximation of the stability coefficients emerging from the full theory, as compared to
 the standard Einstein-Hilbert result.

The ghost-improvement hinges on the new formula \eqref{etaI} for the ghost anomalous dimension $\eta_c$, which is analytic in Newton's constant and completely determined by  $g$ and $\lambda$. At the non-Gaussian fixed point, we obtain $\eta_c^* = -1.77$ for the optimized and $\eta_c^* = -1.85$ for the exponential cutoff. Notably, these numbers are very close to the anomalous dimension of Newton's constant $\eta_N^* = -2$. We believe that this is not accidental, but a consequence of space-time becoming effectively two-dimensional at short distances, where the physics is controlled by the non-Gaussian fixed point \cite{Lauscher:2005qz,Lauscher:2005xz}. The essence of our argument is the observation that the propagator of a field with anomalous dimension $\eta$ is proportional to $p^{-2+\eta}$. Therefore, in the vicinity of the UV fixed point, both the graviton and ghost propagator behave approximately as $p^{-4}$, which translates into a logarithmic correlator in position space. In this case the space-time seen by both fields is effectively two-dimensional, suggesting $\eta_c^* = -2$ in the full theory. A similar spontaneous dimensional reduction has also been observed within the framework of Causal Dynamical Triangulations \cite{Ambjorn:2005db}, and, by now, also in a variety of other quantum gravity approaches \cite{Carlip:2009kf}. It would be very interesting to explore, if this is a first hint pointing at a deeper connection between these seemingly unrelated fields.

\section*{Acknowledgements}
%
We thank D.\ Benedetti, A.\ Codello, J.E.\ Daum, U.\ Harst, P.\ Machado, E.\ Manrique, J.\ Pawlowski, and M.\ Reuter for interesting discussions and A.\ Eichhorn and H.\ Gies for communications on their upcoming work. The research of K.G.\ and F.S.\ is supported by the Deutsche Forschungsgemeinschaft (DFG)
within the Emmy-Noether program (Grant SA/1975 1-1).
%
\begin{appendix}
\section{Heat-kernel techniques}
\label{App:A}
In this appendix we collect the essential formulas for evaluating the operator
traces \eqref{Strace} and \eqref{Gtrace}. We start with the traces containing only Laplacians 
before turning to operator insertions including the background ghost fields in Subsection \ref{A.2}. Their relation
to the threshold-functions employed in the main text is established in Subsection \ref{A.3}. 
\subsection{Early time heat-kernel expansion}
\label{A.1}
The differential operators entering into the traces \eqref{Strace} appear in the form of covariant Laplacians $\Dz = - \Db^2$ only,
and may thus be evaluated using standard early time heat-kernel 
methods for second order differential operators (see \cite{Codello:2008vh} for more details). In this case the heat-kernel expansion takes the form
\be\label{heat1}
\Tr_i \left[ \e^{-s\Delta} \right] = (4\pi s)^{-d/2} \, \int d^dx \sqrt{\gb} \left[ \tr_i \, a_0 + s \, \tr_i \, a_2 + \cdots \right] \, ,
\ee
with coefficients $a_0 = \unit_i$ and $a_2 = \frac{1}{6} \Rb \; \unit_i$ and $\tr_i$ denoting a trace over the internal indices
\be
\tr_{\rm 2T} \, \unit_{\rm 2T} = \half (d+2)(d-1) \, , \qquad \tr_{1} \unit_{1} = d \, , \qquad \tr_{0} \unit_{0} = 1 \, .  
\ee
Here and in the following the dots indicate interaction terms which do not contribute to the truncation.  

By employing a Laplace-transform, eq.\ \eqref{heat1} can be generalized to suitable operator-valued functions $W(z)$,
\be
\Tr_i \left[ W(\Dz) \right] = (4 \pi)^{-d/2} \int d^dx \sqrt{g} \left\{ Q_{d/2}[W] \, \tr_i \, a_0 + Q_{d/2-1}[W] \, \tr_i \, a_2 + \cdots \right\} \, , 
\ee
where
\be\label{Qfct}
Q_n[W] := \int_0^\infty d s s^{-n} \tilde{W}(s) = \frac{1}{\Gamma(n)} \int_0^\infty d z \, z^{n-1} \, W(z) \, .
\ee
Here $\tilde{W}(s)$ is the Laplace transform of $W(z)$ and we used a Mellin-transform to re-express the integral over $\tilde{W}(s)$ in terms of the original function. Employing \eqref{PhiQrel1} to recast the functionals $Q_n$ arising from \eqref{Strace} in terms of the dimensionless threshold functions \eqref{phifunc}, a straightforward computation leads to the right-hand-side of eq.\ \eqref{beta1}.
\subsection{Applying the Anselmi-Benini perturbation method}
\label{A.2}
While the traces $\cS_i$ can be evaluated using the standard heat-kernel techniques for Laplace operators, dealing with the $\cG_i$ is more involved.
Here, the main technical difficulty results from the fact that their arguments contain covariant derivatives which do not organizing themselves into minimal second-order differential operators.
Applying the perturbative heat-kernel technique developed by Anselmi and Benini \cite{Anselmi:2007eq}, this obstruction can be circumvented by the following two-step procedure: In the first step, we use commutator relations to collect all non-minimal operators in a single operator insertion. The second step then uses the heat-kernel formula at non-coincident points to evaluate the operator traces including these ``perturbations''.

To implement the first step, we start from the arguments in \eqref{Gtrace} and express the propagator sandwiched between the $A$'s (or $Q$'s) by its Laplace transform. Then, all the Laplacians reside in an exponential, that can easily be commuted with the $A$'s (or $Q$'s)  
\be\label{comid}
[\cO, \e^{-s \Dz} ] = - s \e^{-s \Dz} [\cO, \Dz] + \cO(s^2) \, .
\ee
Undoing the Laplace-transform leads to the desired form of the trace argument, including only one operator of non-Laplace-type. The prefactor $s$ multiplying the commutator terms thereby translates into a derivative of the original propagator. At the level of the full trace argument, \eqref{comid} implies 
\be
 W(\Dz) Q^{\m\n}{}_{\a} = Q^{\m\n}{}_{\a} W(\Dz) - 2  W^{\prime}(\Dz) \left( D_\sigma Q^{\m\n}{}_{\a} \right) D^\sigma + \cdots \, ,
\ee
and similarly for the other terms.  
Since our sole interest is in isolating the contribution proportional to the ghost kinetic term, all terms containing more than two covariant derivatives acting on $\cb^\m$ and $c_\m$ or involve curvature tensors do not contribute to the truncation and may therefore be dropped. 

After these manipulations, the traces $\cG_i$ assume the generic form
\be\label{conform}
\Tr_i \left[ \, \cO \, W(\Dz)   \right] = \int_0^\infty ds \, \tilde{W}(s) \,  \langle x | \cO \, \e^{-s \Dz} | x \rangle \, .
\ee
Here all Laplacians have been collected in $W(\Dz)$ and $\cO$ symbolizes a (matrix-valued) insertion build from two covariant derivatives acting on the background ghosts together with two (or four) covariant derivatives acting to the right. (See below for an explicit example.) 
The trace \eqref{conform} is evaluated by inserting a complete set of states between the two operators
\be
\begin{split}
 \langle x | \cO \, \e^{-s \Dz} | x \rangle 
= & \; \langle x | \cO | x^\prime \rangle  \langle x^\prime | \e^{-s \Dz} | x \rangle 
=  \, \int d^dx \sqrt{\gb} \, \tr_i \left[ \cO \, H(s, x, x^\prime) \right]_{x^\prime=x} \, .
\end{split}
\ee
Here
\be
H(s, x, x^\prime) := \langle x^\prime | \e^{-s \Dz} | x \rangle = (4 \pi s)^{-d/2} \, \e^{- \frac{\sigma(x,x^\prime)}{2 s}} \, \sum_{n=0}^\infty s^n A_{2n}(x, x^\prime) \, ,
\ee
is the expansion of the heat-function at non-coincident points, with $A_{2n}(x, x^\prime)$ being the off-diagonal heat-kernel coefficients and $\sigma(x,x^\prime)$ denoting half the squared geodesic distance between $x$ and $x^\prime$, satisfying
\be
\sigma(x,x) = 0 \, , \quad \half \sigma^{;\mu} \sigma_{;\mu} = \sigma \, , \quad \sigma_{;\mu\nu}(x,x) = \gb_{\m\n}(x) \, , \quad
\sigma_{;(\alpha_1 \ldots \alpha_n)}(x,x) = 0 \, , \; n \ge 3 \, .
\ee
Notably, $A_{2n}(x, x) = a_{2n}$, so that all but $A_0(x, x)$ give rise to curvature terms. Furthermore, covariant derivatives acting on $A_{2n}(x, x^\prime)$ either vanish or produce higher derivative curvature terms once the coincidence limit is taken \cite{Decanini:2005gt,Anselmi:2007eq}. Thus,
when applying $\cO$ on the function $H(s, x, x^\prime)$, the only terms contributing to the truncation are those where the covariant derivatives act pairwise on $\sigma(x, x^\prime)$.

To illustrate the procedure, let us consider the following explicit example from the $A^\alpha \Aqt_\alpha$ product in $\cG_0$
\be
\begin{split}
\cO_0 = [ \tfrac{1}{d} (\Db_\m c^\a) \Db^\m ] [ -\tfrac{1}{d} (\Db_\n \cb_\a) \Db^\n ] = - \tfrac{1}{d^2} (\Db_\m c^\a) (\Db_\n \cb_\a) \Db^\m \Db^\n + \cdots .
\end{split}
\ee
Here we dropped all terms with more than two covariant derivatives acting on the background ghosts. Substituting $\cO_0$ into \eqref{conform} gives
\be
\begin{split}
\Tr \left[ W(\Dz) \, \cO_0 \,  \right]  & \, = \frac{1}{2 d^2 (4 \pi)^{d/2}} \int_0^\infty ds \, s^{-d/2-1} \, \tilde{W}(s) \int d^dx \sqrt{\gb} (\Db_\m c^\a) (\Db^\m \cb_\a) \\
& \, = \frac{1}{2 d^2 (4 \pi)^{d/2}} \, Q_{d/2+1}[W] \, \int d^dx \sqrt{\gb} \, \cb^\a \Db^2 c_\a \, ,
\end{split}
\ee
where we used \eqref{Qfct} in the second step. Applying this algorithm systematically to all terms in \eqref{Gtrace} 
and subsequently using the relation \eqref{PhiQrel2} then leads to \eqref{Gtrace2}.

\subsection{Threshold functions and cutoff scheme}
\label{A.3}
In order to capture the cutoff-scheme dependence and highlight the structure of the $\beta$-functions it is convenient to express the functionals $Q_n[W]$ in terms of dimensionless threshold functions.
The structures appearing in the $\cG_i$ motivate the definitions
\be\label{phifunc}
\begin{split}
\Phi_{n}^{p,q}(w) :=& \, \frac{1}{\Gamma(n)} \int_0^\infty\!\!d z\; z^{n-1}\frac{R^{(0)}(z)-zR^{(0)\prime}(z)}{ \left( z+R^{(0)}(z)+w \right)^p \, \left(z+R^{(0)}(z)\right)^q} \, , \\
\tphi_{n}^{p,q}(w) :=& \, \frac{1}{\Gamma(n)} \int_0^\infty\!\!d z\; z^{n-1}\frac{R^{(0)}(z)}{\left( z+R^{(0)}(z)+w \right)^p \, \left(z+R^{(0)}(z)\right)^q} \, , \\
\cphi_{n}^{p,q}(w) :=& \,\frac{1}{\Gamma(n)} \int_0^\infty\!\! d z\; z^{n-1}\frac{R^{(0)\prime}(z)(R^{(0)}(z)-zR^{(0)\prime}(z))}{ \left( z+R^{(0)}(z)+w \right)^p \, \left(z+R^{(0)}(z)\right)^q} \, , \\
\hphi_{n}^{p,q}(w) :=& \, \frac{1}{\Gamma(n)} \int_0^\infty\!\!d z\; z^{n-1}\frac{R^{(0)}(z) R^{(0)\prime}(z)}{\left( z+R^{(0)}(z)+w \right)^p \, \left(z+R^{(0)}(z)\right)^q} \, ,
\end{split}
\ee
where the prime denotes the derivative of the shape function $R^{(0)}$, defined below \eqref{IRrule}, with respect to its argument. Observe that
the threshold functions  with $p=0$ do not depend on $w$.
These definitions naturally generalize the ones used in \cite{Reuter:1996cp}, which are recovered as special cases
\be
\begin{split}
\Phi^{p,0}_{n}(w)=\Phi^{p}_{n}(w) \, , \quad
\Phi^{0,q}_{n}(w)=\Phi^{q}_{n}(0) \, , \quad
\tphi^{p,0}_{n}(w)=\tphi^{p}_{n}(w) \, , \quad
\tphi^{0,q}_{n}(w)=\tphi^{q}_{n}(0) \, .
\end{split}
\ee

These threshold functions are closely related to the $Q$-functionals \eqref{Qfct}. 
For the traces \eqref{Strace} we have
\be\label{PhiQrel1}
Q_n\left[ \frac{\p_t (Z^I_k R_k)}{Z^I_k ( P_k + w)^p} \right]
= 2 k^{2(n-p+1)} \left[ \Phi^{p,0}_n(w/k^2) - \half \, \eta_I \, \tphi^{p,0}_n(w/k^2) \right] \, ,
\ee
with the index $I = N,c$ such that the relation holds for both the gravitational and ghost wave-function renormalization.
This relation generalizes to the $Q$-functionals occurring in \eqref{Gtrace}
\be\label{PhiQrel2}
\begin{split}
Q_n[f_1^N] = & \, 2 \, Z^N_k \, k^{2n-4} \left[ 
\Phi^{2,1}_n(-2 \lambda_k) - \half \eta_N \tphi^{2,1}_n(-2\lambda_k) 
\right]\, , \\
Q_n[f_1^c] = & \, 2 \, Z^c_k \, k^{2n-4} \left[ 
\Phi^{1,2}_n(-2 \lambda_k) - \half \eta_c \tphi^{1,2}_n(-2\lambda_k) 
\right]\, , \\
Q_n[f_2^I] = & \, 2 \, Z^I_k \, k^{2n-6}  \left[
\Phi^{2,2}_n(-2\lambda_k) + \cphi^{2,2}_n(-2\lambda_k) - \half \eta_I (\tphi^{2,2}_n(-2 \lambda_k) + \hphi^{2,2}_n(-2 \lambda_k) )
\right] \, .
\end{split}
\ee
In combination with the heat-kernel formulas \eqref{A.3} and \eqref{conform}, these equations allow us to find the $\beta$-functions for $Z_k^N, Z_k^c$ and $\Lambda_k$, \eqref{beta1} and \eqref{dtZc}.

We close this section with a remark on the shape functions employed in the numerical studies of Section \ref{Sect:5}. Unless stated otherwise,
all results are obtained with the optimized cutoff \cite{Litim:2001up}
\be\label{Ropt}
R^{(0) \rm, opt}_k(z) = (1-z) \, \Theta(1-z) \, .
\ee
In this case the integrals appearing in \eqref{phifunc} can be carried out analytically
\be
\begin{split}
\Phi_{n}^{p,q}(w) = \frac{1}{\Gamma(n+1)} \frac{1}{(1+w)^p}  \, , &\qquad
\tphi_{n}^{p,q}(w) = \frac{1}{\Gamma(n+2)} \frac{1}{(1+w)^p} \, , \\
\cphi_n^{p,q}(w) = - \frac{1}{\Gamma(n+1)} \frac{1}{(1+w)^p} \, , &\qquad
\hphi_{n}^{p,q}(w) = - \frac{1}{\Gamma(n+2)} \frac{1}{(1+w)^p} \, .
\end{split}
\ee
Here, the threshold functions degenerate such that they become independent of the index $q$.
When analyzing the cutoff-scheme dependence of our truncation, we also
employ a one-parameter family of smooth exponential shape functions
\be\label{Rexp}
R^{(0) \rm, exp}_k(z; s) = \frac{sz}{\exp(sz)-1} \, . 
\ee
The continuous shape parameter $s$ allows to smoothly vary
the implementation of the IR cutoff. In contrast to the optimized cutoff,
the integrals in the threshold functions cannot be carried out analytically for
this class of cutoffs. Thus we have to resort to numerical integration
when evaluating the threshold functions.

\section{Functions determining the anomalous dimensions}
\label{App:B}
In this appendix, we give the definitions of the functions
$B_i(\lambda)$ and $C_i(\lambda)$, completing the construction
of the anomalous dimensions for the graviton and ghost fields \eqref{etaI}. 
In the following, all threshold functions are evaluated at the argument
$w = -2 \lambda$, which we then suppress for notational simplicity. 
The $B_i$ are exactly the same as those obtained in \cite{Reuter:1996cp}
\be\label{Bdef}
\begin{split}
B_1(\lambda) = & \, \tfrac{1}{3} \, (4\pi)^{1-d/2} \left( d(d+1) \Phi^{1,0}_{d/2-1} - 6d(d-1) \Phi^{2,0}_{d/2} - 4d \Phi^{0,1}_{d/2-1} - 24 \Phi^{0,2}_{d/2} \right) \, , \\
B_2(\lambda) = & \, - \tfrac{1}{6} (4\pi)^{1-d/2} \left( d(d+1) \tphi^{1,0}_{d/2-1} + 6 d(d-1) \tphi^{2,0}_{d/2} \right) \, .
\end{split}
\ee
The quantum corrections from the wave-function renormalization of the ghosts are encoded in 
\be\label{Cdef}
\begin{split}
C_1(\lambda) = & \, (4\pi)^{1-d/2} \left( 2C_{\rm gr}\tphi^{2,1}_{d/2+1} -4d( \tphi^{2,2}_{d/2+2}+\hphi^{2,2}_{d/2+2} ) \right) \, , \\
C_2(\lambda) = & \, (4\pi)^{1-d/2} \left( 2C_{\rm gh}\tphi^{1,2}_{d/2+1} +4d( \tphi^{2,2}_{d/2+2}+\hphi^{2,2}_{d/2+2} ) \right) \, , \\
C_3(\lambda) = & \, \tfrac{1}{3} \, (4\pi)^{1-d/2} \left( 2d \tphi^{0,1}_{d/2-1} +12\tphi^{0,2}_{d/2} \right) \, , \\
C_4(\lambda) = & \, - (4\pi)^{1-d/2} \left( 4C_{\rm gr}\Phi^{2,1}_{d/2+1} +4C_{\rm gh}\Phi^{1,2}_{d/2+1} \right) \, , \\
\end{split}
\ee
with the coefficients $C_{\rm gr}$ and $C_{\rm gh}$ defined in \eqref{Ccdef}.
\end{appendix}

\end{document}